\documentclass[twocolumn]{aastex631}

\usepackage{amsmath}
\usepackage{bm}
\usepackage{diagbox}\usepackage{tabularx}
\usepackage{makecell}

\graphicspath{{./}{Figures/}}

\begin{document}
\title{The Impact of Cometary `impacts' on the Chemistry, Climate, and Spectra of Hot Jupiter Atmospheres}

\author{F. Sainsbury-Martinez and C. Walsh}

\affiliation{School of Physics and Astronomy, University of Leeds, Leeds LS2 9JT, UK}

\begin{abstract} 
Impacts from icy and rocky bodies have helped shape the composition of solar system objects, for example the Earth-Moon system, or the recent impact of comet Shoemaker-Levy 9 with Jupiter. It is likely that such impacts also shape the composition of exoplanetary systems. Here we investigate how cometary impacts might affect the atmospheric composition/chemistry of hot Jupiters, which are prime targets for characterisation. We introduce a parametrised cometary impact model that includes thermal ablation and pressure driven breakup, which we couple with the 1D `radiative-convective' atmospheric model ATMO, including disequilibrium chemistry. We use this model to investigate a wide range of impactor masses and compositions, including those based on observations of Solar System comets, and interstellar ices (with JWST). We find that even a small impactor (R = 2.5 km) can lead to significant short-term changes in the atmospheric chemistry, including a factor $>10$ enhancement in H$_2$O, CO, CO$_2$ abundances, and atmospheric opacity more generally, and the near complete removal of observable hydrocarbons, such as CH$_4$, from the upper atmosphere. 
These effects scale with the change in atmospheric C/O ratio and metallicity. Potentially observable changes are possible for a body that has undergone significant/continuous bombardment, such that the global atmospheric chemistry has been impacted. Our works reveals that cometary impacts can significantly alter or pollute the atmospheric composition/chemistry of hot Jupiters. These changes have the potential to mute/break the proposed link between atmospheric C/O ratio and planet formation location relative to key snowlines in the natal protoplanetary disc.
\end{abstract}

\keywords{Planets and Satellites: Atmospheres --- Planets and Satellites: Composition --- Planets and Satellites: Chemistry --- Planets: HD209458b --- Methods: Computational}

\section{Introduction} \label{sec:introduction}

The impact of icy and rocky bodies has been proposed to have played a significant role in shaping the composition of solar system planets. For example, impacts been invoked to explain Jupiter's super-solar metallicity (see, for example, \citealt{2004jpsm.book...35G,2010SSRv..152..423F}) and stratospheric water abundance \citep{2013A&A...553A..21C}, the dawn-dusk asymmetry of Mercury's exosphere \citep{BENZ1988516,Pokorny_2017}, and the atmosphere of Mars \citep{WOO201987}, as well as shaping Earth's composition through the delivery of complex organic matter to the surface that may have helped to form a habitable Earth \citep[see the review by][]{2020AsBio..20.1121O}. \\

Under the assumption that planet formation proceeds in a similar manner in exoplanetary systems, as has happened in the solar system, one would expect bombardment and impacts to have also shaped the composition and atmospheres of exoplanets. This is particularly likely for so-called hot Jupiters, with many/most of the formation pathways for these objects requiring that they start life in the outer disc, beyond the snowlines where at least H$_2$O solidifies. Then they have been proposed to migrate inwards towards their observed radii (with orbital periods $<$ 10 days), very close to their host star (Type II and III migration, \citealt{2007prpl.conf..655P,Chatterjee_2008,2012ARA&A..50..211K}). This process is likely to be highly disruptive, imparting significant angular momentum into the system and hence scattering many objects out of their orbits \citep{2003AdSpR..31.2563M}, and potentially into the gravitational well of the migrating hot Jupiter (see \citealt{2014IAUS..310..194R} for a review of how Jupiter's migration might have shaped our solar system). {Recent observations reveal several young  planets which may fulfil these requirements, having either migrated to short orbital radii whilst the systems debris disk is extant, or showing signs of significant post-migration accretion. Examples include AU Mic b/c \citep{aumicb,2021A&A...649A.177M}, AF Lep b \citep{2023AJ....166..198Z}, K2-33b \citep{2016Natur.534..658D}, and V1298 Tau b/c/d/e \citep{2019ApJ...885L..12D,2022NatAs...6..232S}. }  \\

Using our own solar system as a basis, modelling of the migration of gaseous planets by \citet{2023arXiv230707089B} has shown that as much as 1\% of the material in the proto-Kuiper-belt (which likely contained around 30 Earth masses of material) might have impacted Jupiter due to the migration of Neptune. This is equivalent to the mass of the entire upper and mid atmosphere (P $<$ 10 bar) of a Jupiter-like planet, suggesting that, even with material settling over long-time-scales, post formation material delivery might have had a significant impact on the composition of Jupiter-like planets.\\
Post migration cometary impacts might also play a role in changing the atmospheric chemistry of hot Jupiters. For example, \citet{2023PSJ.....4..139N} suggest that a cometary impact with radius R $>$ 1.0 km  should occur every $\simeq$ 150 years with Jupiter. Gravitational focusing and the increased orbital velocity associated with short orbits should increase this rate for more massive hot Jupiters. For example, a cometary impact rate of 0.01 yr$^{-1}$, assuming R $=$ 2.5 km icy come, would be enough to drive a 0.1\% change in the mass of the outer atmosphere of HD209458b over its entire lifetime. 
And we have evidence that comets can survive close enough to the host star to impact a hot Jupiter, with \citet{2017ApJ...845...27N} suggesting that comets with radii $>$ 5 km can survive thousands of approaches, as well as the potential detection, via a shallow transit, of a 2.5 km radius comet orbiting HD 172555 at a distance of $0.05\pm0.01$ AU \citep{2023A&A...671A..25K}. Similar objects have also been observed in our own solar system, and are typically referred to as sun-approaching/grazing comets, with some objects even being observed to impact the sun and be destroyed \citep{Sekanina_2005,comets_review}. \\
These incoming comets do not necessarily have to share the same formation location as the object that they impact, instead they can form throughout the disk \citep{2003AdSpR..31.2563M}. Hence significant differences in composition, and hence carbon-to-oxygen (C/O) ratios, might also occur \citep{2017MNRAS.467.2845A}. Consequently, incoming cometary material might lead to significant changes in atmospheric chemistry and temperature. For single impacts these changes are likely to be localised and somewhat time-limited, as was the case for the Shoemaker–Levy 9 (SL9) impact with Jupiter \citep{doi:10.1126/science.263.5148.787,1995A&A...295L..35G,1995ApJ...438..957F,2005A&A...434..343A,2006ApJ...646..642K,TOTH2006162,2016Icar..264....1F}. Whereas if the hot Jupiter has experienced significant historical bombardment, either due to a period of heavy bombardment or a historically high impact rate, the changes might be global and long-lasting, potentially breaking the proposed link between atmospheric C/O ratio and planet formation location relative to key snowlines in the natal protoplanetary disc. \\

Here we investigate such a scenario, modelling the effects of individual cometary impacts on a localised region of the atmosphere of an HD209458b-like exoplanet. At early times we treat our models as being representative 
of the localised effects of a cometary impact on a near-terminator region of the atmosphere, with this region being chosen due to its potential(!) observability via transit spectroscopy, as well as the potential for photochemistry to play a major role in the chemistry, something we expect to distinguish impacts with hot Jupiters from cooler Jupiter-like objects. On the other hand, later times, due to to the neglect of horizontal mixing in our approach, are treated as being representative of a saturated state in which the hot Jupiter has undergoing significant bombardment by cometary material with a fixed C/O ratio (composition), resulting in changes to the global chemistry and composition of the atmosphere. \\

In \autoref{sec:method}, we introduce our model, which couples a cometary ablation and break-up model with the 1D `radiative-convective' atmospheric model, ATMO, that includes disequilibrium chemistry effects. This includes a discussion of the cometary properties considered here, including the radius, density, tensile strength, and composition. Then, in \autoref{sec:results}, we analyse select results from our ensemble of cometary impact models in more detail. This includes examples of how individual impacts might affect the atmospheric chemistry, composition, and temperature, and in turn how these changes might affect observable features, such as transmission spectra (assuming most optimistic cases). We also explore how the arrival of cometary material impacts the abundances of molecules in the upper atmosphere, linking these changes to changes in the atmospheric C/O ratio and metallicity. We finish, in \autoref{sec:discuss}, with some concluding remarks, discussing the implications of our results as well as potential plans for future studies which will extend this work beyond the 1D regime. \\

\section{Method} \label{sec:method}
To understand how cometary impacts affect the atmospheres of hot Jupiters, we have coupled a parametrised cometary impact model (\autoref{sec:impact_model}), which models ablation at low pressures and break-up at high pressures, with the radiative-convective atmospheric disequilibrium chemistry model, ATMO (\autoref{sec:ATMO}). This coupled model requires a number of inputs, including planetary parameters, an unperturbed atmospheric model, and cometary parameters, such as the radius, density, tensile strength and composition of the impacting comet (\autoref{sec:comet_param}).

\subsection{Cometary Impact Model} \label{sec:impact_model}
For the sake of computational efficiency and broader model comparability, our cometary ablation and breakup model is highly parametrised. We assume that the comet encounters the atmosphere at a zero angle of incidence (i.e. $\cos(\theta)=1$) and remains undeformed (i.e. spherical) until breakup. Further, whilst the impact itself is modelled in a time-dependent manner, tracing the journey of the comet through the atmosphere, the resulting changes to the atmospheric chemistry via mass deposition are all applied to the atmospheric model concurrently. \\
The impact itself can be split into two phases: in the upper atmosphere, where the pressure is low, the comet slows due to atmospheric drag, drag whose friction drives surface ablation. However, as the density increases, so to does the drag, leading to an increase in the ram pressure and eventual breakup when the ram pressure exceeds the tensile strength of the comet \citep{PASSEY1980211,2016ApJ...832...41M}. \\
During the ablation phase, at each time-step, the velocity of the impactor is calculated using the velocity evolution equation of \citet{PASSEY1980211} (and references therein): 
\begin{align}
    \frac{dV}{dt} = g - \frac{C_{D}\rho_{a}AV^{2}}{M}, \label{eq:velocity_evo} 
\end{align}
where
\begin{align}
    A=S_{F}\left(\frac{M}{\rho_{c}}\right)^{\frac{2}{3}}, \label{eq:shape_factor} 
\end{align}
is the effective cross-sectional area of the comet, $g$ is the gravitational acceleration of the planet, $C_{D}$ is the drag coefficient, $\rho_{a}$ is the atmospheric density, $S_{F}$ is the shape factor of the comet ($S_{F}=1.3$ for a sphere), and $\rho_{c}$, $V$, and $M$ are the cometary density, velocity, and mass respectively. \\
Similarly, the ablation driven mass deposition is given by:
\begin{align}
    \frac{dM}{dt} = -\frac{C_{H}\rho_{a}AV^{2}}{2Q}\left(\frac{V^{2}-V_{cr}^{2}}{V^{2}}\right), \label{eq:mass_evo} 
\end{align}
where $C_{H}$ is the heat transfer coefficient, $Q$ is the heat of ablation of the cometary material (see \autoref{tab:comet_parameters}), and $V_{cr}=3$ km s$^{-1}$ is the critical velocity below which no ablation occurs due to the reduced deposition of frictional energy \citep{PASSEY1980211}.  \\
At the same time, the ram pressure ($P_{ram}$) is also calculated and checked against the tensile strength ($\sigma_{T}$) of the comet, with breakup occurring when:
\begin{align}
    P_{ram} &> \sigma_{T},\\
    P_{ram} &= C_{D}\rho_{A}V^{2}. \label{eq:ram}
\end{align}
At this point the impact is terminated and any remaining mass is distributed deeper into the atmosphere over a pressure scale height using an exponentially decaying function in order to simulate the the rapid breakup (i.e. shredding/crushing) of the comet and the resulting mass distribution due to the inertia of the incoming material \citep{2006ApJ...646..642K}. \\
Whilst relatively simple, when properly configured (see \autoref{sec:comet_param}), this approach reproduces both the observed breakup location of comet Shoemaker-Levy 9 with Jupiter in 1994 (around 1 bar - see, for example, \citealt{2006ApJ...646..642K}), as well as breakup locations calculated using more complex physical processes which are believed to underlie cometary destruction. This includes cometary deformation (i.e. pancaking - e.g. \citealt{1994ApJ...434L..33M,1995ApJ...438..957F,2016ApJ...832...41M}) as well as instabilities such as the Rayleigh-Taylor \citep{1995ApJ...438..957F,2005ApJ...626L..57A} or Kelvin-Helmholtz \citep{KORYCANSKY20021} instabilities. The latter of which drives breakup by inducing surface waves which lead to the shedding of droplets of surface material, and eventually grow to be on the order of the size of the comet, disrupting it, and leading to breakup. \\
Next, once the mass distribution profile for the impact has been calculated, it is then converted into a mass density profile by evenly distributing the mass at each vertical level over a 150 km sided box. This box size was chosen to ensure that comets of all radii of interest can be modelled: bigger box sizes lead to smaller comets having a negligible impact on the atmospheric composition, whereas smaller box sizes lead to instabilities in the chemical solver as the most-massive comets completely overwhelm the unperturbed initial atmosphere. This occurs when any constituent component of the atmosphere becomes more fractionally abundant than hydrogen, even temporarily (e.g. during initial material deposition) - this can easily be avoided by ensuring that the incoming mass is distributed/mixed over a sufficiently large area. {Note that, due to the 1D nature of our atmospheric models, changing the box size is degenerate with changing the comets mass (increasing the box size means distributing the mass over a larger area, resulting in local density changes that are equivalent to a smaller comet spread over the original, smaller, area). Hence we chose to keep the box sized fixed for our models, varying the comet radius/density as a more intuitive mechanism by which to adjust the amount of cometary material delivered to the atmosphere.}  \\
Finally the cometary composition (\autoref{tab:comet_models}) is used to convert this mass density into a number density for each of the constituent species, which is combined with our unperturbed hot Jupiter model to generate an initial molecular abundance profile that will be evolved by ATMO. \\

The impact of a comet/planetesimal might also lead to atmospheric heating due to at least some of the kinetic energy lost by the impacting comet being transferred to the surrounding atmosphere, with the exact ratio of kinetic energy lost to ablation and the surrounding atmosphere being controlled by the heat transfer coefficient $C_{H}$ \citep{2005A&A...434..343A,2018A&A...618A..19R}. For our models, we set $C_{H}=0.5$ (which is the upper limit of the range of potential values given by \citet{1995Icar..116..131S} and \citet{2005A&A...434..343A}), evenly splitting the lost kinetic energy between the comet (and hence thermal ablation) and atmospheric heating (which takes the form $0.5C_{H}\left(m_{t}v_{t}^2 - m_{t+1}v_{t+1}^2\right) = \frac{3}{2}k_{B}T$). Here we found that the long-term effect of such a heating term was minimal. Whilst it does lead to some initial heating of the upper atmosphere, the short radiative time-scales of this region paired with the strong day-side irradiation caused by the very close host star mean that, after a few days/weeks of model time, models with and without kinetic energy driven heating are near indistinguishable. This remains true on the dark nightside, when redistribution is efficient and hence strong day-night heating can play a similar role \citep{2020SSRv..216..139S}. However, for the sake of energy conservation, we retain this initial heating term for all models shown here. 

\subsection{ATMO and HD209458b} \label{sec:ATMO}

ATMO is a 1D `radiative-convective' disequilibrium chemistry model for simulating the structure and composition of hydrogen-dominated planetary atmospheres, such as hot Jupiters and brown Dwarfs \citep{2014A&A...564A..59A,2015ApJ...804L..17T,2016ApJ...817L..19T,2016A&A...594A..69D}.  It has been regularly applied to interpret transit observations, both as a forward and retrieval (i.e. recovering observations) model, including the analysis of early-release JWST observations (e.g. \citealt{2022ApJ...940L..35F}). \\ 
In fully-consistent modelling mode, ATMO iteratively solves for both the atmospheric pressure-temperature profile and chemical abundances whilst including radiative transport \citep{2014A&A...564A..59A,2016A&A...595A..36A} and chemical kinetics \citep{2016A&A...594A..69D}, including vertical mixing and photochemistry. The inclusion of chemical kinetics is key for a number of reasons, ranging from the strong vertical mixing associated with molecular abundance gradients that result from the impact, to the significant fraction of a comet's mass which is made up of water, a significant opacity source \citep{2000ApJ...537..916S,2008ApJ...678.1419F,2010ApJ...719..341B} that is susceptible to photodissociation in the presence of strong UV radiation \citep{2019ApJ...871..158F}. Further, a number of recent studies, such as \citet{2014RSPTA.37230073M} and \citet{2020A&A...635A..31M} have shown that the inclusion of chemical kinetics, 
turbulent and molecular diffusion, photochemical processes, as well as free election interactions can all significantly influence the resulting steady-state structure. \\

For this study, we consider the archetypal hot Jupiter HD209458b as our test bed and initial unperturbed atmosphere.
HD209458b has a radius of 1.35 R$_{\mathrm{J}}$, surface gravity $\log(g)=2.98$, and orbital radius of 0.048 AU \citep{2000ApJ...529L..45C,10.1111/j.1365-2966.2010.17231.x}. The host star is sun-like (a G0 star vs the G5 sun) with R$=$1.114 R$_{\odot}$, hence the short-wavelength stellar stellar spectrum is a modification of the sun-like spectrum of \citet{1970SAOSR.309.....K}\footnote{\url{http://kurucz.harvard.edu/stars.html}}\citep[and][]{1991ASIC..341..441K,2003IAUS..210P.A20C}, whilst the XUV spectrum (used for 34 reactions included in the photochemistry scheme) is purely sun-like since HD209458 exhibits sun-like activity levels \citep{10.1093/mnras/stw2421}. This irradiation intercepts our 1D atmospheric model at an angle of 45$^\circ$, chosen because such a configuration typically provides a good fit to observations \citep{oro73551}. \\
Vertically, our models extends between $10^{-5}$ and $10^2$ bar, allowing us to model both the upper atmosphere, which is probed by transmission spectra, as well as the deep, optically thick, atmosphere in which the cometary break-up and resulting inertia-driven mass deposition occurs. These regions are linked via vertical mixing, which is parametrised using a constant eddy diffusion coefficient $K_{zz}=10^{9}$ cm$^2$s$^{-1}$, a value that was chosen to be broadly compatible with prior studies using ATMO \citep{2016A&A...594A..69D,Venot_2020} and of hot Jupiter atmospheres \citep{2011ApJ...737...15M,2014ApJ...780..166M,2020A&A...635A..31M,2023A&A...669A.142S,2023ApJ...946L...6M}. Note that small changes in the adopted value of the mixing strength ($K_{zz}=10^{8}\rightarrow10^{10}$) did not significantly alter our results, other than the time taken for the model to settle. \\
The radiative transfer scheme takes a correlated-k approach \citep{GOODY1989539,2014A&A...564A..59A} to computing the total opacity, with the model including 20 opacity sources: H$_2$-H$_2$ collision induced absorption or CIA, H$_2$-He CIA, H$_2$O, CO, CO$_2$, CH$_4$, NH$_3$, Na, K, L, Rb, Cs, FeH, PH$_3$, H$_2$S, HCN, C$_2$H$_2$, SO2, Fe, H$^-$ free-free and bound-free - see table 1 of \citealt{2020A&A...637A..38P} for details of these opacity sources. 
Under this approach, the incoming radiation is split in a number of bands in which opacities are more easily parametrised, 32 for our actual atmospheric models and 500 when calculating transmission-spectra/optical depths, uniformly distributed in wavenumber space between 31 cm$^{-1}$ and 50000 cm$^{-1}$. \\
Finally, the chemical-kinetics scheme used is the $C_0$-$C_2$ chemical network of \citet{2012A&A...546A..43V}, which includes 957 reversible and 6 irreversible reactions involving 105 species, plus helium as a potential third body in some reactions. 

\begin{table}
\centering
\begin{tabular}{lr}
Parameter & Value \\ \hline
Radius $R$ (km) & $2.5\rightarrow30$ \\
Box Size (km$^2$) & $150\times150$ \\
Density $\rho_c$ (g cm$^{-3}$) & $1\rightarrow2$ \\
Initial Velocity $V_{0}$ (km s$^{-1}$)  & 20  \\
Heat Transfer Coefficient $C_{H}$  & 0.5 \\
Drag Coefficient $C_{D}$ & 0.5 \\
Latent Heat of Ablation  Q(erg g$^{-1}$) & $2.5\times10^{10}$ \\
Tensile Strength $\sigma_{T}$ (erg cm$^{-2}$) & $4\times10^{6}$
\end{tabular}
\caption{Parameter (ranges) for the impacting comets considered in this work.}
\label{tab:comet_parameters}
\end{table}
\begin{table*}
\centering
\begin{tabular}{l|llllllll|l|l}
 & H$_2$O & CO & CO$_2$ & CH$_4$ & NH$_3$ & CH$_3$OH & NCO & HCN & C/O & Label \\ \hline
Solar Nebula \citep{Lodders_2003} & 1.0 & 0.0 & 0.0 & 0.65 & 0.18 & 0.0 & 0.0 & 0.0 & 0.65 & A  \\
Comet - ver. A \citep{2004come.book..391B} & 1.0 & 0.1 & 0.05 & 0.1 & 0.01 & 0.0 & 0.0 & 0.0 & 0.21  & B  \\
Comet - ver. B \citep{2015AA...583A...1L} & 1.0 & 0.06 & 0.19 & 0.0 & 0.0 & 0.0 & 0.0 & 0.0 & 0.17  & C \\
High CO$_2$ Comet - Modified Comet ver. B & 0.06 & 0.19 & 1.0 & 0.0 & 0.0 & 0.0 & 0.0 & 0.0 & 0.48  & D  \\
Protoplanetary Disk \citep{2005ApJ...622..463P} & 1.0 & 0.99 & 0.32 & 0.035 & 0.1 & 0.1 & 0.0 & 0.53 & 0.53  & E  \\
Planetesimals - ver. A \citep{2005SSRv..116...77A} & 1.0 & 0.15 & {0.37} & 0.01 & 0.05 & 0.04 & 0.0 & 0.0 & {0.27}  & F  \\
Planetesimals - ver. B \citep{2005SSRv..116...77A} & 1.0 & 0.15 & {0.37} & 0.01 & 0.006 & 0.04 & 0.0 & 0.0 & {0.27} & G  \\
JWST ICE - ver. A \citep{2023NatAs...7..431M} & 1.0 & 0.44 & 0.2 & 0.026 & 0.126 & 0.15 & 0.003 & 0.0087 & {0.42}  & H  \\
JWST ICE - ver. B \citep{2023NatAs...7..431M} & 1.0 & 0.28 & 0.13 & 0.019 & 0.107 & 0.07 & 0.002 & 0.0067 & {0.32} & I  
\end{tabular}
\caption{Molecular abundances for the 9 different cometary abundances considered here. All abundances are given with respect to the most abundant molecule, which is H$_2$O for the vast majority of compositions. The only exception is the modified high CO$_2$ cometary composition, which was run in order to expand the number of models at higher C/O.}
\label{tab:comet_models}
\end{table*}
\subsection{Cometary Parameters} \label{sec:comet_param}

Cometary parameters (\autoref{tab:comet_parameters}) have been selected to be broadly compatible with observations of comets within our own solar system, as well as recent observations of interstellar ices, presuming that said ices are preserved during the star and planet formation process. Where a choice existed in the exact value of a parameter, such as tensile strength, either an aggregate of reported values was selected, or we ran a series of tests in order to find that value that best reproduced the breakup of comet Shoemaker-Levy 9 (SL9). \\
To that end, we consider:
\begin{itemize}
    \itemsep0em 
    \item A cometary radius range of between 2.5 km and 30 km (see the review by \citealt{2011ARA&A..49..281A}, including references therein, as well as \citealt{2015Icar..260...60S}).
    \item Two cometary densities, 1 g cm$^{-3}$ (slightly compacted or `dirty' H$_2$O ice) and 2 g cm$^{-3}$ (a denser compacted ice mix), both of which are broadly compatible with models/observations (see the review of \citealt{2008M&PS...43.1033W}) under the assumption that the atmospheric pressure has resulted in compressive waves that enhance sub 1 g cm$^{-3}$ densities to $\sim1$ g cm$^{-3}$ \citep{1995ApJ...438..957F}.
    \item An initial cometary velocity of 20 km s$^{-1}$, in agreement with the majority of the models of \citet{KORYCANSKY20021}.
    \item A drag coefficient equal to that of a sphere in a moderately turbulent flow, $C_D=0.5$.
    \item A latent heat of ablation equivalent to that of pure water, i.e. $Q=2.5\times10^{10}$ erg g$^{-1}$ \citep{1995ApJ...438..957F,2016ApJ...832...41M}. 
\end{itemize}
As for the tensile strength, estimates for comets vary widely, with studies of both different bodies and different locations on the same body reporting significantly different values. For example, \citet{doi:10.1126/science.aab0464} (and \citealt{MOHLMANN2018251,2016SoSyR..50..225B}) report values for comet 67P/Churyumov–Gerasimenko, based upon Rosetta mission observations, ranging from $6.5\times10^{4}$ erg cm$^{-2}$ at the Philae lander impact site, to $4\times10^{7}$ erg cm$^{-2}$ within the Abydos valley region of the comet. As such, during initial model development, we tested cometary impact models with tensile strengths between that of loosely packed snow and solid water ice, $10^{3}$ and $10^{7}$ erg cm$^{-2}$ respectively \citep{Petrovic2003ReviewMP}, exploring which value best reproduced the breakup pressure of comet SL9 ($0.1\rightarrow1$ bar). We found that a tensile strength of between $10^{6}$ and $10^{7}$ erg cm$^{-2}$ was optimal, and as such we settled on $\sigma_{T}=4\times10^{6}$ erg cm$^{-2}$, as used by \citet{2016ApJ...832...41M}. \\ 

Finally, we come to the cometary composition (\autoref{tab:comet_models}), which we treat a bit differently. Rather than converging on a single composition based upon a broad mean of literature values, we instead investigate nine different composition models. This allows us to study how differences in the composition of the incoming material are reflected in the final atmospheric temperature and chemistry. The compositions we consider include: 
\begin{itemize}
    \itemsep0em 
    \item A model of the solar nebula that lead to the formation of our solar system (\citealt{Lodders_2003} - table 11).
    \item A solar system cometary composition model taken from (\citealt{2004come.book..391B} - referred to comet ver. A).
    \item A model of the composition of comet 67P/Churyumov-Gerasimenko based upon Rosetta mission data taken from \citet{2015AA...583A...1L} (referred to comet ver. B).
    \item A modification of the aforementioned 67P cometary composition in which the CO$_2$ abundance has been significantly enhanced (note that we assume that all other cometary parameters, such as Q, are unchanged).
    \item A model of the composition of the protoplanetary disk CRBR 2422.8-3423, based upon Spitzer observations, from \citet{2005ApJ...622..463P}.
    \item A pair of planetesimal models, based upon the observationally driven planet formation models of \citet{2005SSRv..116...77A}, for which two different N$_2$/NH$_3$ ratios where considered (N$_2$/NH$_3$ = 1 - ver. A -  and N$_2$/NH$_3$=10 - ver. B).
    \item A pair of interstellar ice models based upon JWST observations of two sight-lines (ver. A and ver. B) probing the very outer envelope of the class 0 protostar Cha MMS1, taken from \citet{2023NatAs...7..431M}.
\end{itemize}
Note that, in general, our composition models are carbon depleted relative to the sun (C/O=0.55 - \citealt{2009ARA&A..47..481A}), suggesting that the observed objects/regions either have a lower C/O ratio than our solar-system, formed interior to the snowlines of key carbon bearing species (which significantly affects cometary C/O ratios - \citealt{2011ApJ...743L..16O}), or contain additional refractory carbon that is missing from the listed volatile inventories.

\begin{figure}[tbp] %
\begin{centering}
\includegraphics[width=0.85\columnwidth]{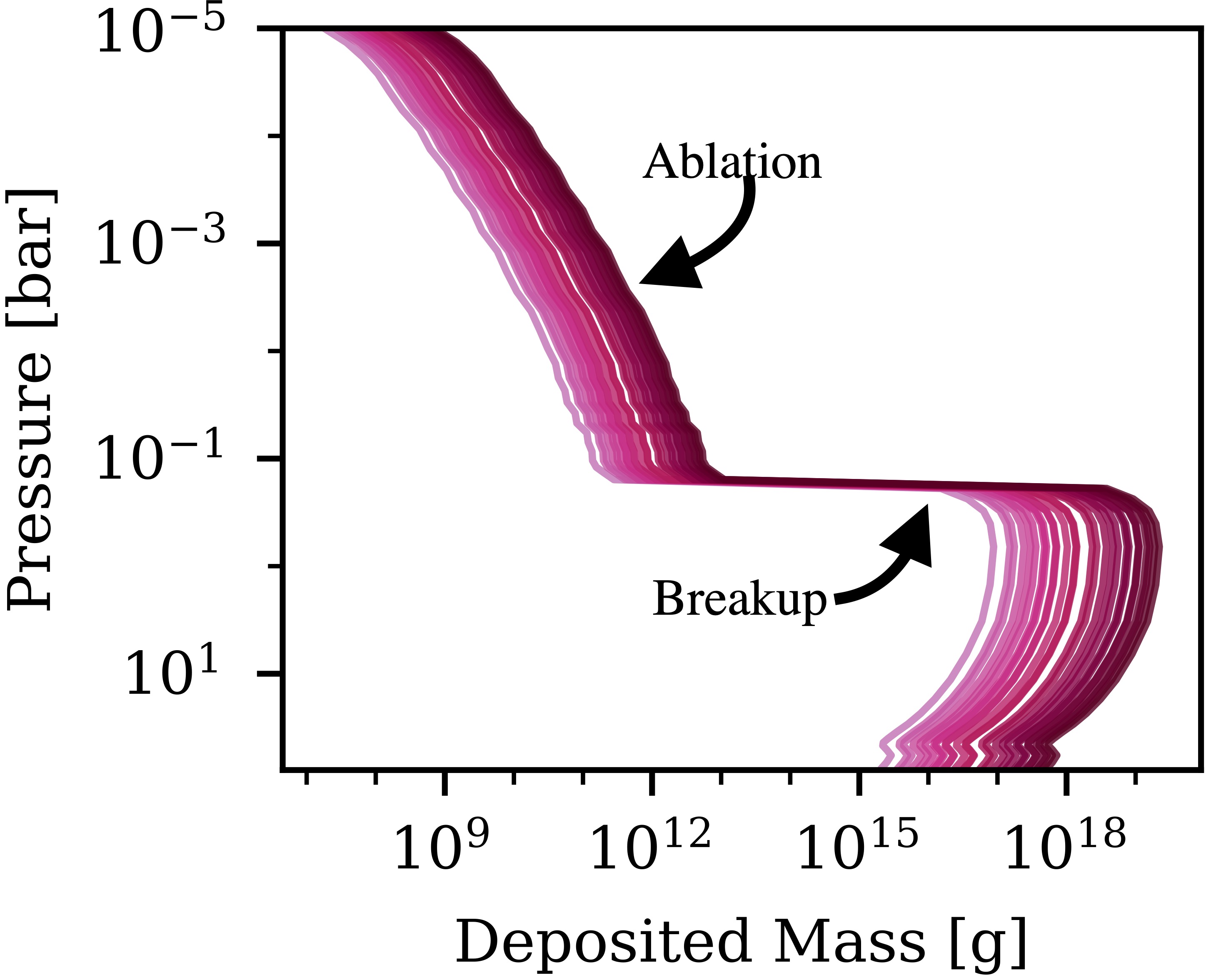}
\caption{ Initial mass distribution profiles generated by our cometary ablation and break-up model for comets with radii between 2.5 km and 30 km and densities of either 1 or 2 g cm$^{-3}$. Note that darker colours indicate higher cometary masses, and that the mass distribution profiles are independent of cometary composition.      \label{fig:mass_deposition} }
\end{centering}
\end{figure}
\begin{figure*}
\gridline{\fig{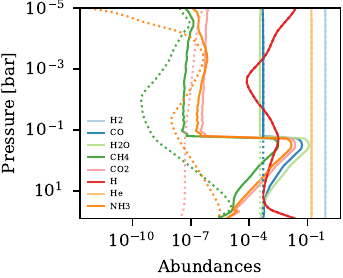}{0.45\textwidth}{\normalsize A) $R =$ 5 km comet starting composition}
          \fig{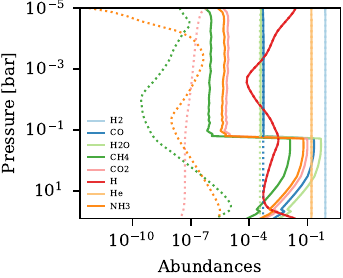}{0.45\textwidth}{\normalsize B) $R =$ 22.5 km comet starting composition}}
\gridline{\fig{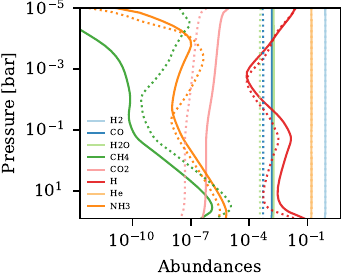}{0.45\textwidth}{\normalsize C) $R =$ 5 km comet steady-state}
          \fig{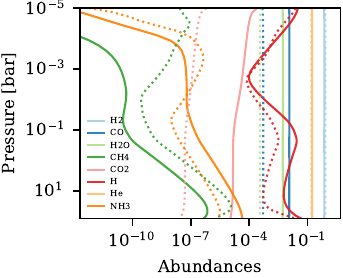}{0.45\textwidth}{\normalsize D) $R =$ 22.5 km comet steady-state}}
\caption{Initial (top) and steady-state (bottom) atmospheric abundance {(i.e. volume mixing ratio)} profiles (solid lines) for two JWST Ice ver. A composition cometary composition impact models, one with $R =$ 5 km (left) and the other with $R =$ 22.5 km (right). Here we focus on 8 key molecules which are either highly abundant, or which are likely to play an outsized role in either the atmospheric chemistry or as an observable atmospheric feature. Further, in order to aid in our analysis and in comparisons between models, we also include the atmospheric abundance profiles from the unperturbed HD20458b model (dashed lines). Note that at each pressure level, the abundances are given relative to the total atmospheric composition.   \label{fig:JWST_ICE_A_Start_finish_ab}}
\end{figure*}
\section{Results} \label{sec:results}

To investigate the possible effects that cometary impacts might have on the atmospheric chemistry of hot Jupiters, such as HD209458b, we have run over 400 cometary impact models to steady-state, distributed between the nine different compositions under consideration (\autoref{tab:comet_models}). Here we will focus most of our analysis on a single composition which reflects our best view to date of ices in an interstellar cloud: the JWST observations of the NIR38 sight-line through the cloud near the class 0 protostar Cha MMS1 (see JWST Ice ver. A in \autoref{tab:comet_models} and \citealt{2023NatAs...7..431M}), returning to the ensemble of cometary compositions in \autoref{sec:atmospheric_co_metal}. \\

\subsection{The Initial Mass Deposition Profile}

We start by exploring the initial mass-deposition profiles (\autoref{fig:mass_deposition}) that are common to all of our cometary models regardless of composition. Here, the two distinct regimes that make up the cometary impact are clear: in the upper atmosphere, where the atmospheric density/pressure is low, the incoming comet retains its global integrity, with only surface material being ablated and deposited into the atmosphere at a rate that depends upon the size of the comet (\autoref{eq:mass_evo}) - larger comets exhibit higher ablation rates. However, as the comet continues its downwards journey, the atmospheric density/pressure increases, increasing both the rate of ablation as well as the stresses on the comet. Eventually the density becomes high enough that the ram pressure exceeds the tensile strength of the comet (\autoref{eq:ram}), leading to breakup and the final distribution of any remaining cometary material. As can be seen in \autoref{fig:mass_deposition}, the exact location at which the breakup occurs is dependent upon the cometary mass, with larger comets experiencing weaker breaking due to atmospheric drag (\autoref{eq:shape_factor}) and hence moving faster, increasing the ram pressure for an otherwise identical atmospheric density. \\

\subsection{Example Initial and Steady-State Atmospheres} \label{sec:steady-state}

Examples of the initial fractional molecular abundance {(i.e. volume mixing ratio)} profiles that result from processing the mass deposition profile with a cometary composition model (\autoref{sec:method}) and combining it with a unperturbed HD209458b-like atmospheric model can be found on the top row of \autoref{fig:JWST_ICE_A_Start_finish_ab}, with the left profile resulting from a $R=5$ km, $\rho=1$ g cm$^{-3}$, cometary impact (one of the smaller comets considered here) and the right from a $R=22.5$ km, $\rho=1$ g cm$^{-3}$, cometary impact (one of the larger comets).  \\
Here we can see a number of features which will influence the steady-state atmospheric composition. For example, because the abundances of CO$_2$, CH$_4$, and NH$_3$ in the unperturbed atmosphere are low, even the smallest amount of deposited cometary material will lead to a large increase in the initial abundance, with the relative increase being much larger than that found for H$_2$O or CO. But this is not to say that the initial enhancement in H$_2$O and CO abundances is small:  The $R=22.5$ km comet impact model reveals that at around the breakup pressure, H$_2$O has become the second most abundant component of the atmosphere, and CO is more abundant than He. However all of this is temporary: vertical mixing leads to this strong local enhancement rapidly becoming more uniformly spread throughout the atmosphere, with H$_2$O/CO becoming less abundant than He at all pressures in around three months of simulation time. \\

Moving onto the steady-state fractional abundances, which are shown on the bottom row of \autoref{fig:JWST_ICE_A_Start_finish_ab}, we find that the impact of the smaller comet (left) has had a significant effect on the atmospheric chemistry. For example, the water content of the atmosphere is over an order of magnitude higher than that found in the unperturbed atmosphere, with this effect growing to a factor of $\sim1000$ as the impactor size increases (right). \\
Starting with the smaller  $R=5$ km cometary impact, which reaches steady-state after approximately 25 years of simulation time, we find that H$_2$O, CO, and CO$_2$ have been uniformly enhanced, with H$_2$O overtaking CO in overall abundance and CO$_2$ showing the strongest relative enhancement: 80 times versus four and two times for H$_2$O/CO respectively. This can be linked to the change in C/O ratio, and more generally to the overall oxygen content, of the atmosphere - for example the unperturbed HD209458b atmosphere's C/O ratio was 0.58 (i.e. close to solar), whereas this atmosphere's C/O ratio is 0.48. Consequently, we have seen a shift in the carriers of carbon away from hydrocarbons, such as CH$_4$ or C$_2$H$_2$ (reduced by, on average a factor of $10^{4}$ and $10^{9}$ respectively), and towards oxygen rich atmospheric components, such as CO$_2$. This occurs despite the fact that additional CH$_4$ was introduced to the atmosphere by the cometary impact. Looking into CH$_4$ in more detail, we find that the largest reduction in abundance (five orders of magnitude) occurs in the upper atmosphere, at pressures less than $10^{-2}$ bar, suggesting that the primary driving force behind this change is photolysis, specifically the photolysis of water.  During and after the cometary impact, ablation and vertical mixing result in a high abundance of water in the very upper atmosphere. Here the strong stellar irradiation causes water photodissociation, freeing hydrogen, oxygen, and hydroxide (OH) that goes on to form CO, CO$_2$, O$_2$, NO, NO$_2$ etc. (see the extended \autoref{fig:JWST_ICE_A_multi_comp}). This explanation is reinforced by the abundance of excited oxygen, O($^1$D), which is massively enhanced at early times {(average abundance of $1.3\times10^{-8}$ vs an unperturbed abundance of $4.4\times10^{-13}$, for our 22.5 km impact case)} when H$_2$O photolysis is at its peak, and remains enhanced by {around two} orders of magnitude at steady state {($4.02\times10^{-11}$)}. It may also be responsible for the factor $\sim10$ enrichment of atomic hydrogen (H) found in the mid to deep atmosphere (although the temperature of the deep atmosphere may also play a role, see \autoref{sec:PT}). Note that we have confirmed that the primary driving forces behind this CH$_4$ destruction are reactions with H and OH, both are which are created by water photolysis in the upper atmosphere. \\

Moving on to the $R=22.5$ km cometary impact, which reaches steady-state after approximately 45 years of simulation time, we find an atmosphere that is highly influenced by the low C/O ratio and hence high oxygen content of the incoming material. Once again we find that H$_2$O, CO, and CO$_2$ have been uniformly enhanced, by 14, 22, and $\sim1000$ times respectively. However unlike in the $R=5$ km cometary impact, we no longer find that H$_2$O is more abundant than CO. Despite the fact that more H$_2$O is delivered to the atmosphere than CO by the comet, the final CO/H$_2$O ratio tilts even more in favour of CO than in the unperturbed model (see \autoref{sec:non_linear_water}). However, just because the oxygen is not in water does not mean that it is not present, as the C/O ratio will attest: here we find C/O = 0.42, suggesting that the carbon-to-oxygen ratio of the atmosphere is converging towards that of the incoming comet (this can be seen in \autoref{sec:atmospheric_co_metal}/\autoref{fig:full_comp_co}). The oxygen-rich nature of the atmosphere is also reflected in the abundances of hydrocarbons, such as CH$_4$, whose global fractional abundance has dropped by a factor of $\sim10^{4}$ and very upper atmosphere abundance has dropped by more than ten orders of magnitude. We once again attribute this effect to water photolysis, which, as we discuss in \autoref{sec:non_linear_water}, has grown non-linearly with incoming cometary mass.\\

In the above analysis we have not discussed how nitrogen bearing molecules behave in detail. This is because, unlike the molecules discussed above, we find neither a uniform increase or decrease in abundance for NH$_3$, NO, HCN or even oxygen-rich NO$_2$. Instead we find that the change in abundance is pressure dependent. For example, at lower impactor masses, we find that NH$_3$ is enhanced in the upper and lower atmosphere whilst being diminished in the mid-atmosphere (between $\sim10^{-4}$ and $\sim10^{-2}$ bar). Then, as the impactor mass increases, we find a general increase in NH$_3$ except in the very upper atmosphere ($P<\sim10^{-4}$ bar) where the trend is reversed and NH$_3$ is further diminished. Yet, despite this increase, for all the models considered here, the abundance of NH$_3$ in the mid-atmosphere never exceeds that found in our unperturbed HD209458b model. \\
As for why this trend occurs, the general enhancement in NH$_3$, and other nitrogen bearing molecules, is due to a combination of the influx of nitrogen from the impacting comet as well as the increase in atmospheric metallicity favouring the formation of heavier molecules (see \autoref{sec:atmospheric_co_metal} for more details). At the same time, the upper atmosphere suppression can be linked to photolysis, both directly, by breaking down NH$_3$, and indirectly, via the presence of additional `free' oxygen linked to water photolysis. Together, these effects also lead to the formation of more oxygen-rich nitrogen-bearing molecules, such as NO and NO$_2$, both of which are enhanced by over five orders of magnitude, and both of which are heavy, and hence are prone to sinking out of the upper atmosphere. 
\begin{figure}[tbp] %
\begin{centering}
\includegraphics[width=0.90\columnwidth]{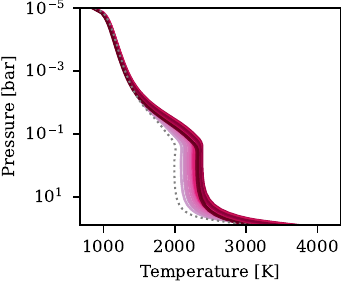}
\caption{ Temperature-Pressure profiles for only those cometary impact models with JWST Ice ver. A composition (\autoref{sec:comet_param}), impactor radii of between 2.5 km and 30.0 km, and densities of either 1 or 2 g cm$^{-3}$. Here, darker colours indicate more massive comets, and the unperturbed HD209458b P-T profile is shown in grey. At very high impactor masses, the temperature of the deep atmosphere is reduced with respect to intermediate-mass impactors.  
    \label{fig:temperature_distribution_JWST_ICE_A} }
\end{centering}
\end{figure}

\subsection{How Cometary Impacts Affect The Pressure-Temperature Profile} \label{sec:PT}

Our cometary impacts affect more than just the atmospheric chemistry. As shown in \autoref{fig:temperature_distribution_JWST_ICE_A}, the temperature of the atmosphere has increased at all pressures, with the strongest heating occurring in the mid/deep atmosphere (P$>10^{-1}$ bar). Here we find that, except for the largest impactors, the temperature of the deep atmosphere increases with impactor size (we discuss why this relationship breaks-down for large impactors in \autoref{sec:non_linear_water}). This change is driven by the presence of water: essentially, water acts as a greenhouse gas, reducing the amount of heat radiated away from the deep atmosphere and hence increasing the temperature. As for where the energy to heat the deep atmosphere comes from, ATMO includes an artificial internal energy source which is designed to heat the deep atmosphere in order to account for the radius-inflation of hot Jupiters, something which 1D `radiative-convective' convective models cannot account for a priori (this is the so-called radius-inflation problem). However the artificial nature of this heating in ATMO does not preclude such an effect from occurring in a real hot Jupiter that experiences a similar H$_2$O enrichment. As discussed in \citet{2017ApJ...841...30T} and \citet{2019A&A...632A.114S,2021A&A...656A.128S,2023MNRAS.tmp.1835S}, the most likely solution to the radius-inflation problem is the vertical transport of heat from the upper to deep atmosphere, where it heats the internal adiabat (and also causes the internal adiabat to form at lower pressures, increasing the apparent radius). This heating would be similarly affected by an enhanced water greenhouse effect since the input energy source does not rely on radiative transport, and deep cooling - radiative loss - would be similarly suppressed. As such, given the link between the temperature of the deep atmosphere and the impactor mass, and hence atmospheric C/O ratio (and metallicity), we propose that said C/O ratio might act as a secondary effect controlling the exact level of radius inflation observed for a hot Jupiter. \\
Note that the change in temperature of the lower/deep atmosphere has consequences for the amount of free/atomic hydrogen in the deep atmosphere, with heating causing increasing amounts of molecular hydrogen to dissociate into atomic hydrogen. 

\begin{figure}[tbp]
\begin{centering}
\includegraphics[width=0.90\columnwidth]{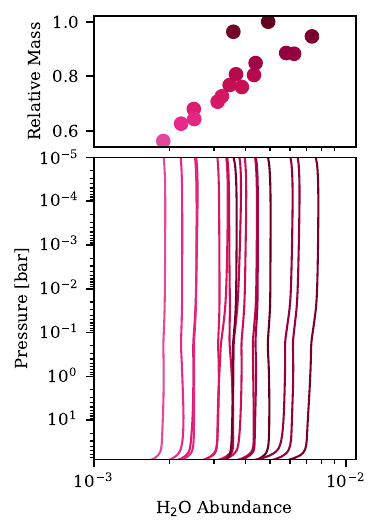}
\caption{ Mass-Abundance (top) and Pressure-Abundance (bottom) curves detailing how the fractional abundance of H$_2$O varies with cometary impactor size. Here we only consider models with the JWST Ice ver. A cometary composition, and have split out the mass-abundance profile (top) in order to demonstrate the non-linear relationship between cometary mass and steady-state H$_2$O abundance.  \label{fig:JWST_ICE_A_H2O_comp_advanced} }
\end{centering}
\end{figure}

\subsection{The Non-Linear Relationship Between Comet Size and Water Abundance} \label{sec:non_linear_water}

As discussed above, and shown in \autoref{fig:JWST_ICE_A_H2O_comp_advanced}, we find that the abundance of H$_2$O departs from a linear cometary-mass/H$_2$O-abundance relationship at very high impactor masses. More specifically we find that there is a threshold for water deposition above which the steady-state water abundance is either constant or lower than that found with a lower impactor mass. Note that the exact location of this threshold is composition dependent (i.e., cometary water fraction dependent): for example we find that the water enhancement starts to be suppressed when $R>20$ km for a 1 g cm$^{-3}$ comet with JWST Ice ver. A composition. \\
Our analysis suggests that the cause of this H$_2$O deficiency at high impactor mass is a combination of vertical mixing and photolysis. Massive impacts cause a correspondingly massive initial influx of H$_2$O, primarily in the lower atmosphere, which is then rapidly mixed into the upper atmosphere due to the resulting density/molecular gradient, where it then undergoes photolysis. Whilst this mixing/photolysis occurs for all impacts, as the impactor mass grows, so too does the amount of water that ends up in the upper atmosphere at early times. This water is then exposed to significant UV irradiation, leading to strong initial photolysis and hence an overall reduction in the total total atmospheric H$_2$O by the time that the atmosphere has reached steady state. This conclusion is reinforced by {both} our alternate (higher/lower) $K_{zz}$ tests, which reveals that more efficient diffusion of water out of the upper atmosphere mutes this enhanced photolysis, {and our multi-impact tests (in which cometary material is spread over 10 smaller impacts), which reveals slightly slower photolysis and hence higher steady-state water abundances (due to the slower delivery of water to the atmosphere)}. \\
Interestingly, we do not find such a break in the cometary-mass/abundance relation for other molecules. However, theoretically, it should occur for any other molecule that undergoes photolysis, when its initial abundance is large enough. For example hints of this trend are found for our high CO$_2$ cometary composition models, where we find that the rate of CO$_2$ abundance increase with impactor mass appears to slow at large impactor sizes. However because of the numerous other paths by which CO$_2$ can become depleted in these models, the effect is rather muted. 

\begin{figure*}
\gridline{\fig{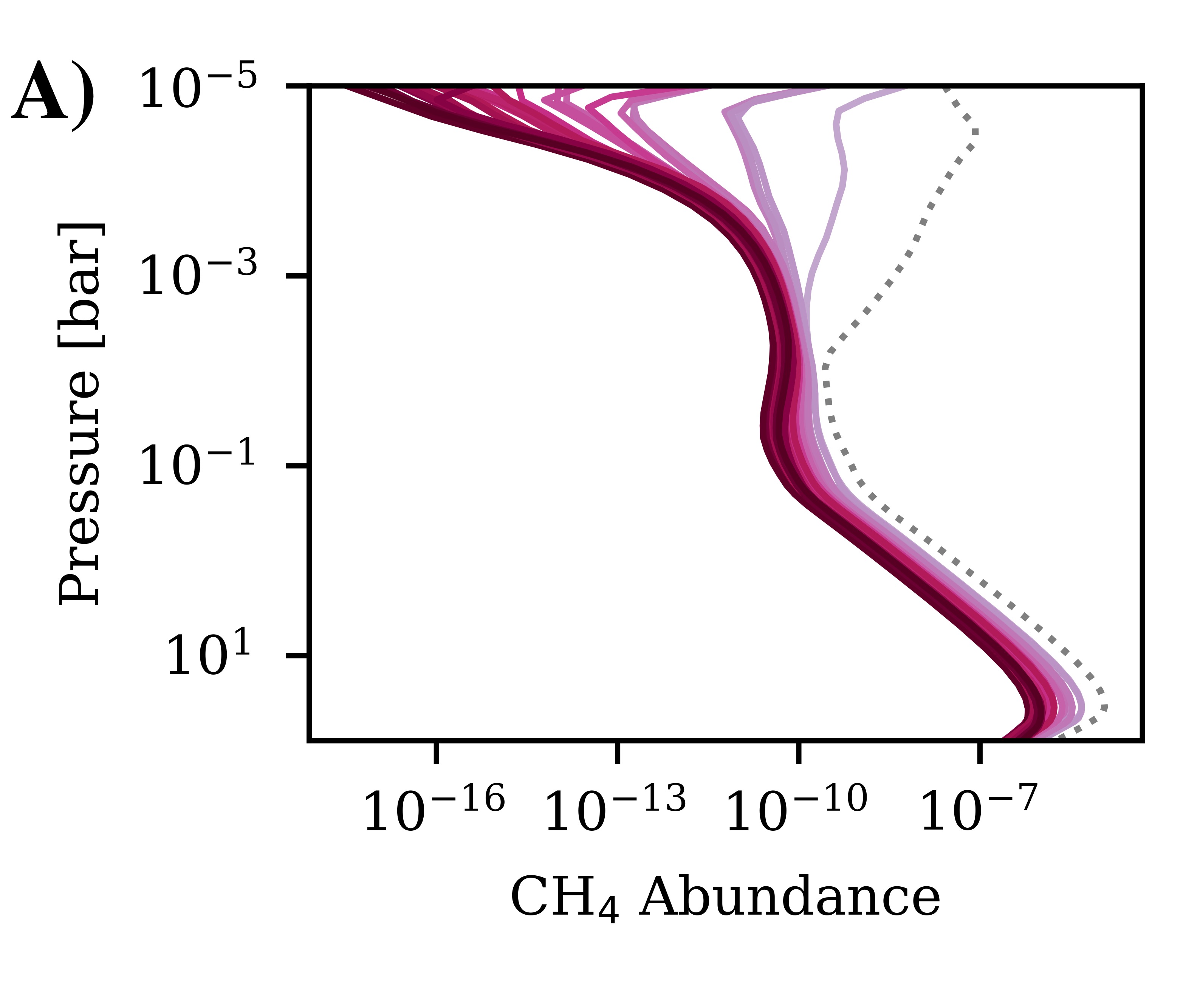}{0.45\textwidth}{}
          \fig{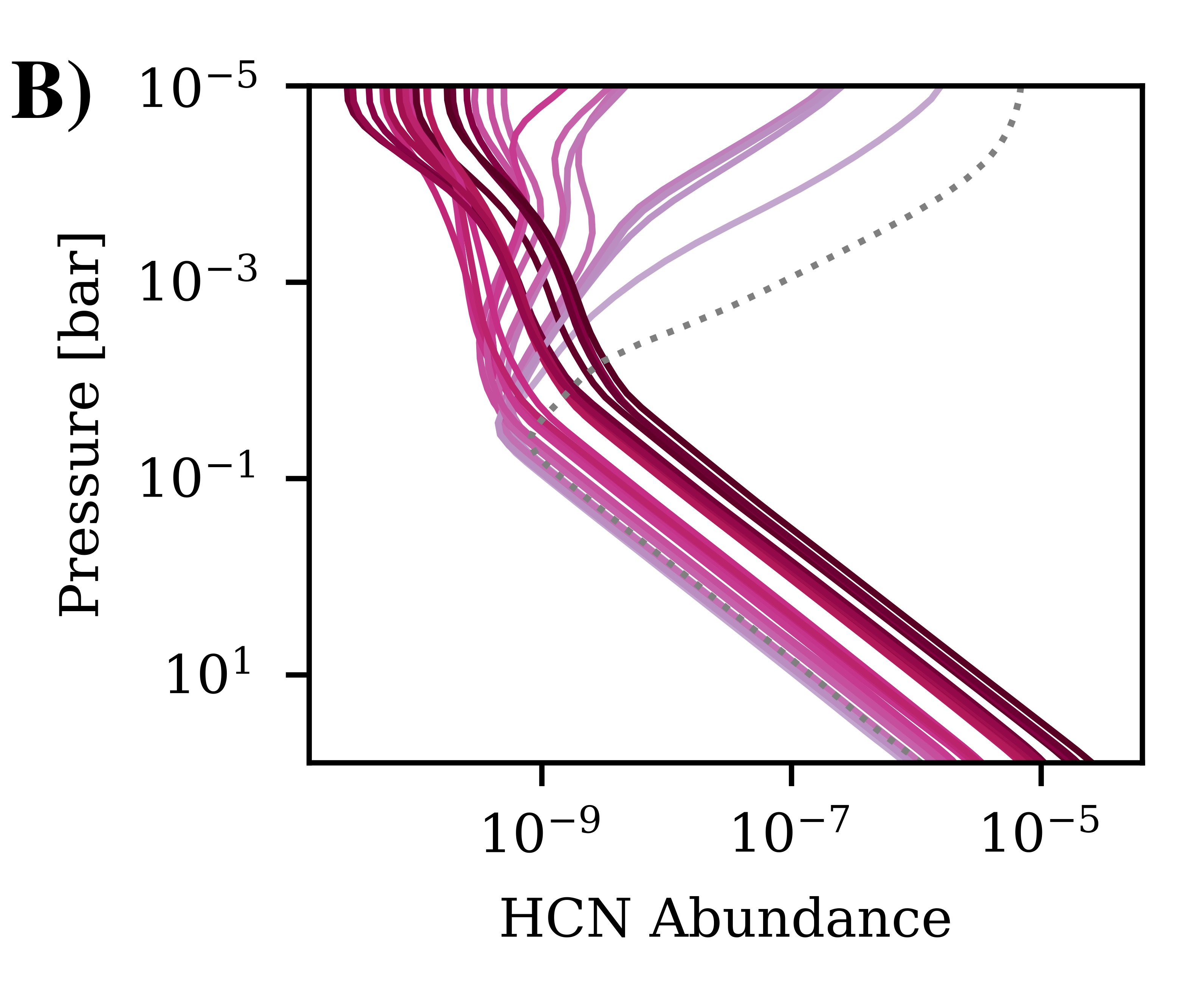}{0.45\textwidth}{}}
\gridline{\fig{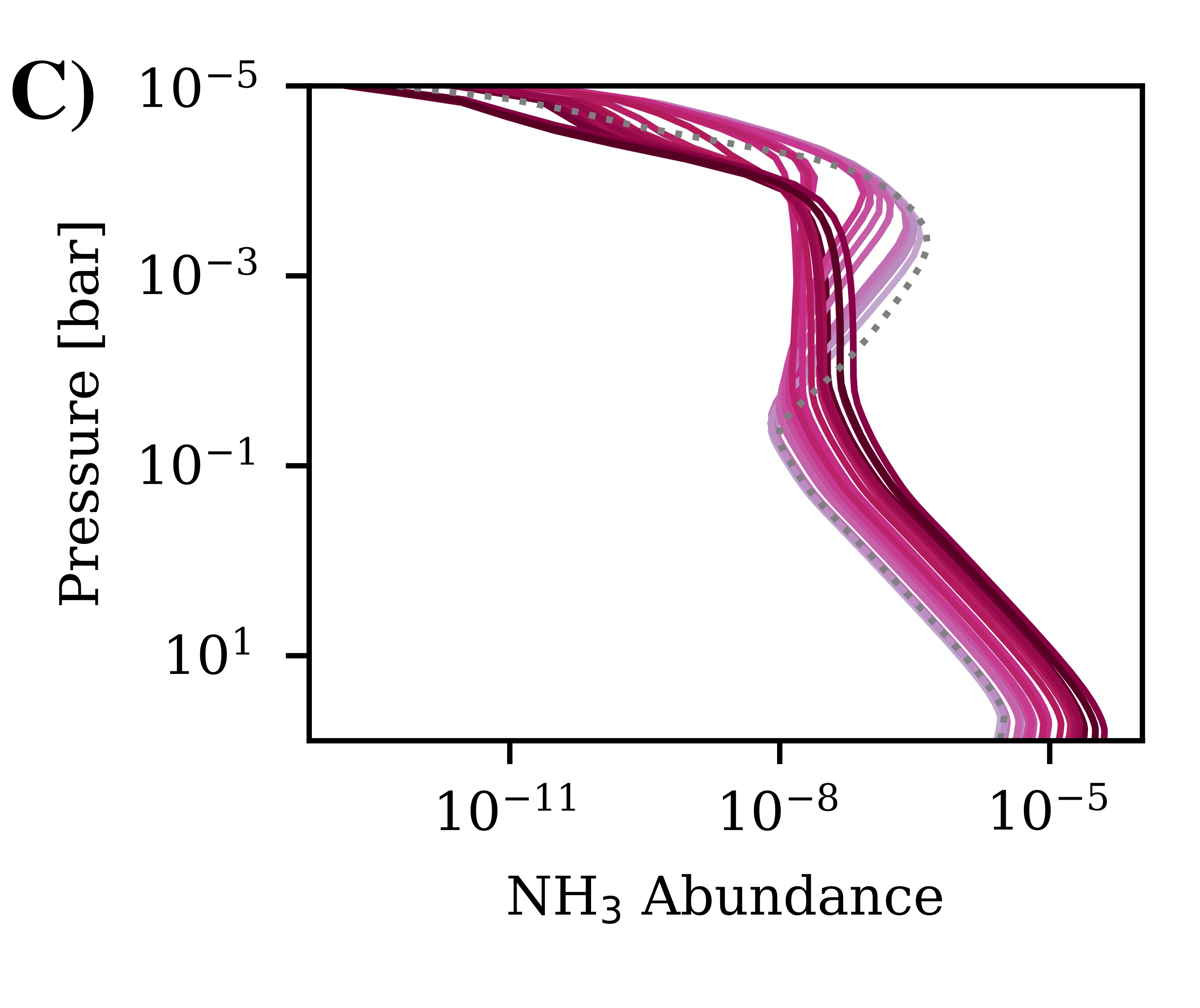}{0.45\textwidth}{}
          \fig{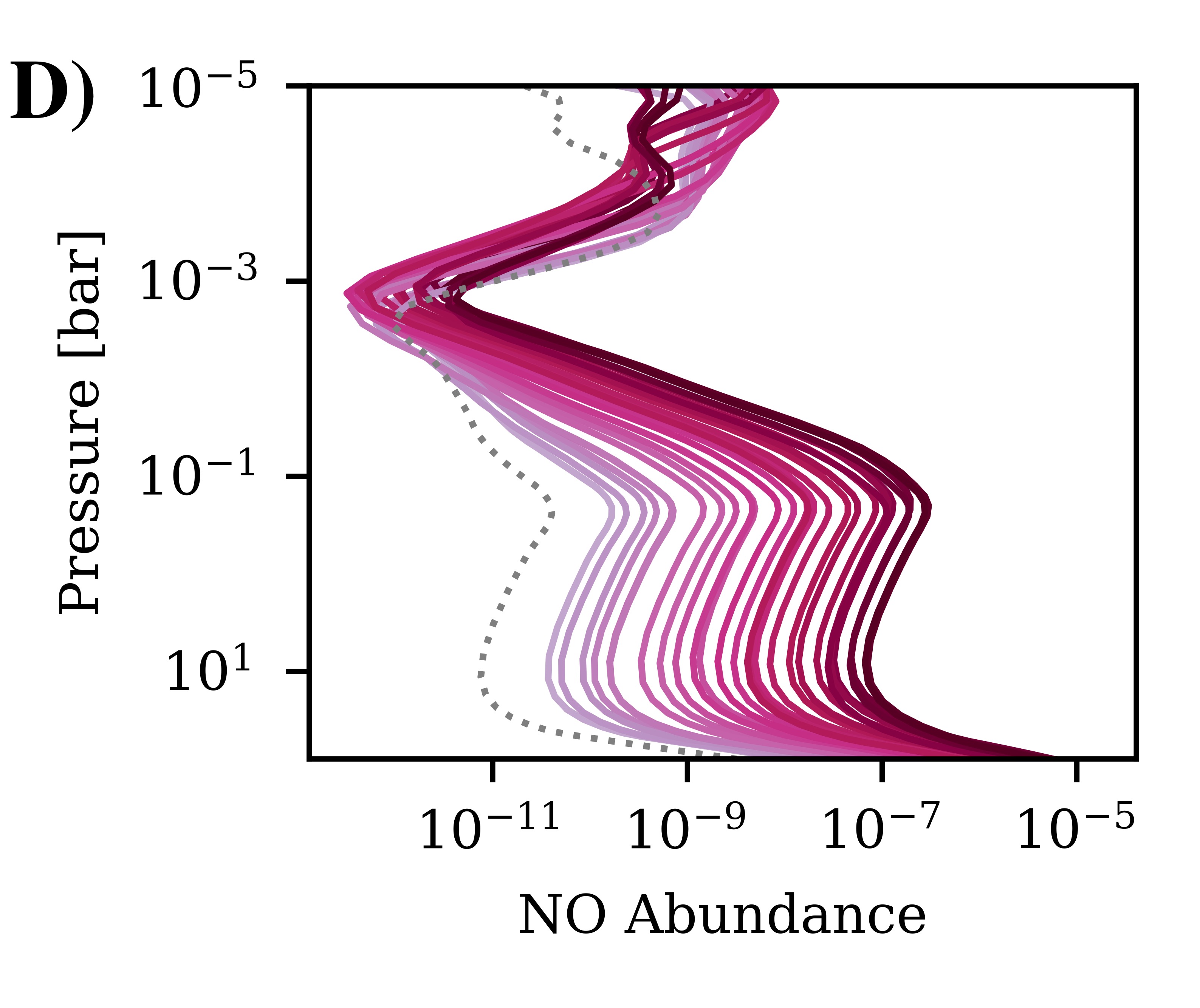}{0.45\textwidth}{}}
\caption{Pressure-Abundance curves for four molecules, CH$_4$ (A), HCN (B), NH$_4$ (C), and NO (D), detailing how cometary impactor size influences steady-state abundances. Here we focus on impact models with JWST Ice ver. A compositions, densities of 1 or 2 g cm$^{-3}$, and cometary radii which vary between 2.5 km and 30 km, using the colour of each abundance curve to denote the relative mass of the impacting comet, with darker colours indicating more massive objetcs. Further, as a point of reference, we also include the relevant abundance curves from our unperturbed HD209458b model as a dashed grey line. Note that abundance curves for all potentially observable molecules are given in the appendix.  \label{fig:JWST_ICE_A_multi_comp}}
\end{figure*}

\subsection{Comet Size and Molecular Abundance Trends} \label{sec:size}

We next explore how the abundances of other molecules change with impactor size (mass/radius). Here we find that our models typically fall into one of a number of regimes, examples of which are shown in \autoref{fig:JWST_ICE_A_multi_comp}. Note that the below trends are discussed relative to an R $=$ 2.5 km, $\rho=$ 1 g cm$^{-3}$, cometary impact model, not the unperturbed HD209458b-like atmosphere. Consequently, a trend may be referred to as an enhancement with impactor mass/radius even when the molecular abundance is less than that found in the unperturbed model - it is just less by a smaller amount. \\

The first regime is a simple, uniform, increase in abundance with impactor mass, as seen for H$_2$O below the `saturation' limit (\autoref{fig:JWST_ICE_A_H2O_comp_advanced}). Other molecules that behave like this include other oxygen bearing molecules, such as CO, CO$_2$, and O$_2$. \\
Next we have molecules such as NO (\autoref{fig:JWST_ICE_A_multi_comp}D), O($^3$P), H,  and OH, which follow a similar trend of increasing abundance with impactor mass, but with a pressure dependence in the strength of said enhancement: generally we find that the relative enhancement increases with pressure. We attribute this pressure dependence to both the effects of photochemistry which is confined to the upper atmosphere, as well as molecular mass. \\
We also find a large number of molecules which are increasingly diminished in abundance with impactor mass in the upper atmosphere, but which show much weaker impactor mass-dependence in the lower atmosphere. An example of this trend can be seen for CH$_4$ (\autoref{fig:JWST_ICE_A_multi_comp}A), which, as we discuss in \autoref{sec:steady-state}, shows a relatively weak decrease in abundance with impactor mass in the deep atmosphere, and a much stronger dependence of abundance on impactor mass in the upper atmosphere, where free oxygen and hydroxide is abundant, leading to oxygen-rich molecules being chemically preferred. A similar story holds true for other carbon/nitrogen-rich and oxygen-poor molecules, such as NH$_3$ (\autoref{fig:JWST_ICE_A_multi_comp}C - although here the deep enhancement is stronger due to overall nitrogen enrichment by the comet), CH, CH$_3$ and C$_2$H$_4$. \\
Finally, there are a number of molecules for which no clear trend is evident. For example, HCN (\autoref{fig:JWST_ICE_A_multi_comp}B) exhibits a rather interesting impactor-size/abundance dependence in the upper atmosphere: when the impactor is relatively small, we find that the abundance of HCN decreases with increasing impactor size, however when the impactor size is relatively large, the trend is reversed, and the abundance instead increases slightly with impactor size.  The underlying cause of this relationship is likely due to the non-linear interaction of a number of competing effects, including the overall enrichment in atmospheric nitrogen caused by larger impactor, a more general increase in atmospheric metallicity with increasing cometary mass deposition, and photochemistry in the upper atmosphere changing the local atmospheric composition. For HCN, in the lower atmosphere, the primary controller of the abundance is the total nitrogen content of the atmosphere, hence we find little to no enhancement in HCN for any cometary impacts that do not carry nitrogen-bearing molecules, such as the Comet ver. A composition. Furthermore, even for those impacts that do introduce additional nitrogen to the atmosphere, the upper atmosphere HCN abundance still remains less than that found in our unperturbed model. \\
Examples of additional molecules, which may be abundant enough in the upper atmosphere to impact observations, are shown in extended \autoref{fig:JWST_ICE_A_multi_comp}. Together, these profiles reveal the complex ways in which molecular abundances change as we adjust the composition of the atmosphere, reinforcing the need for robust models when interpreting observations of planetary atmospheres.

\begin{figure*}
\gridline{\fig{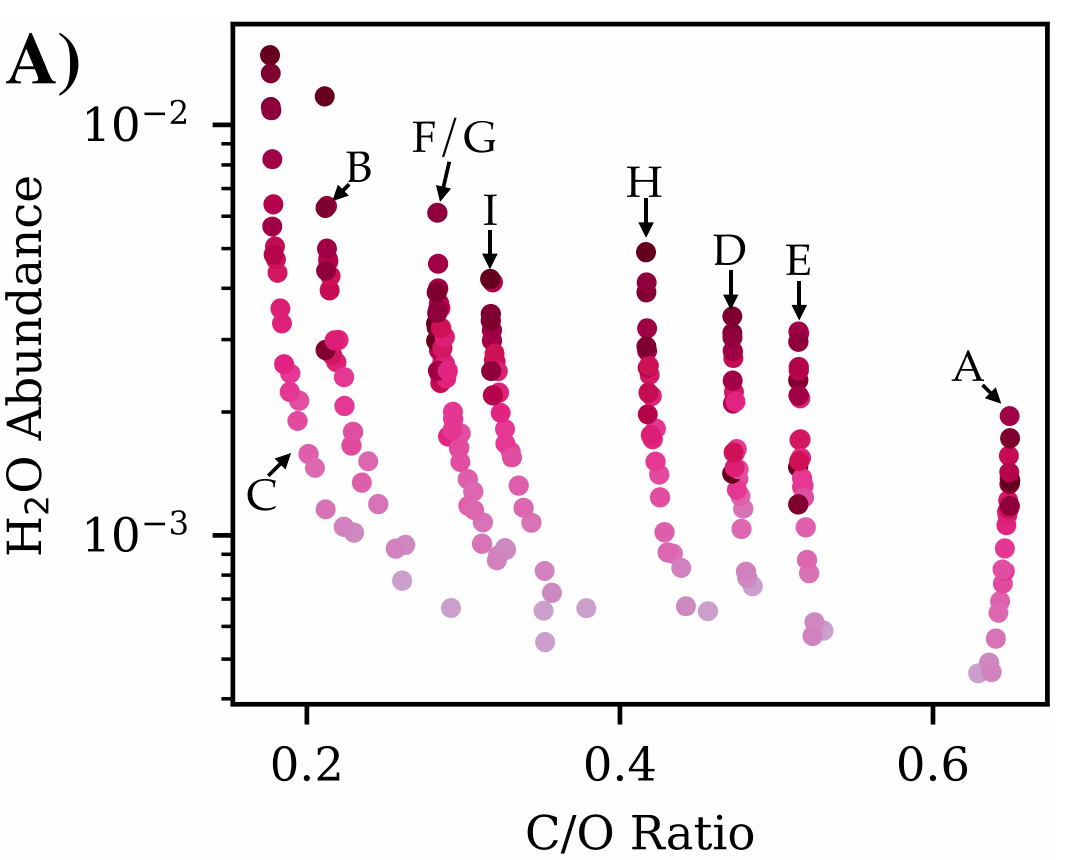}{0.75\columnwidth}{}
          \fig{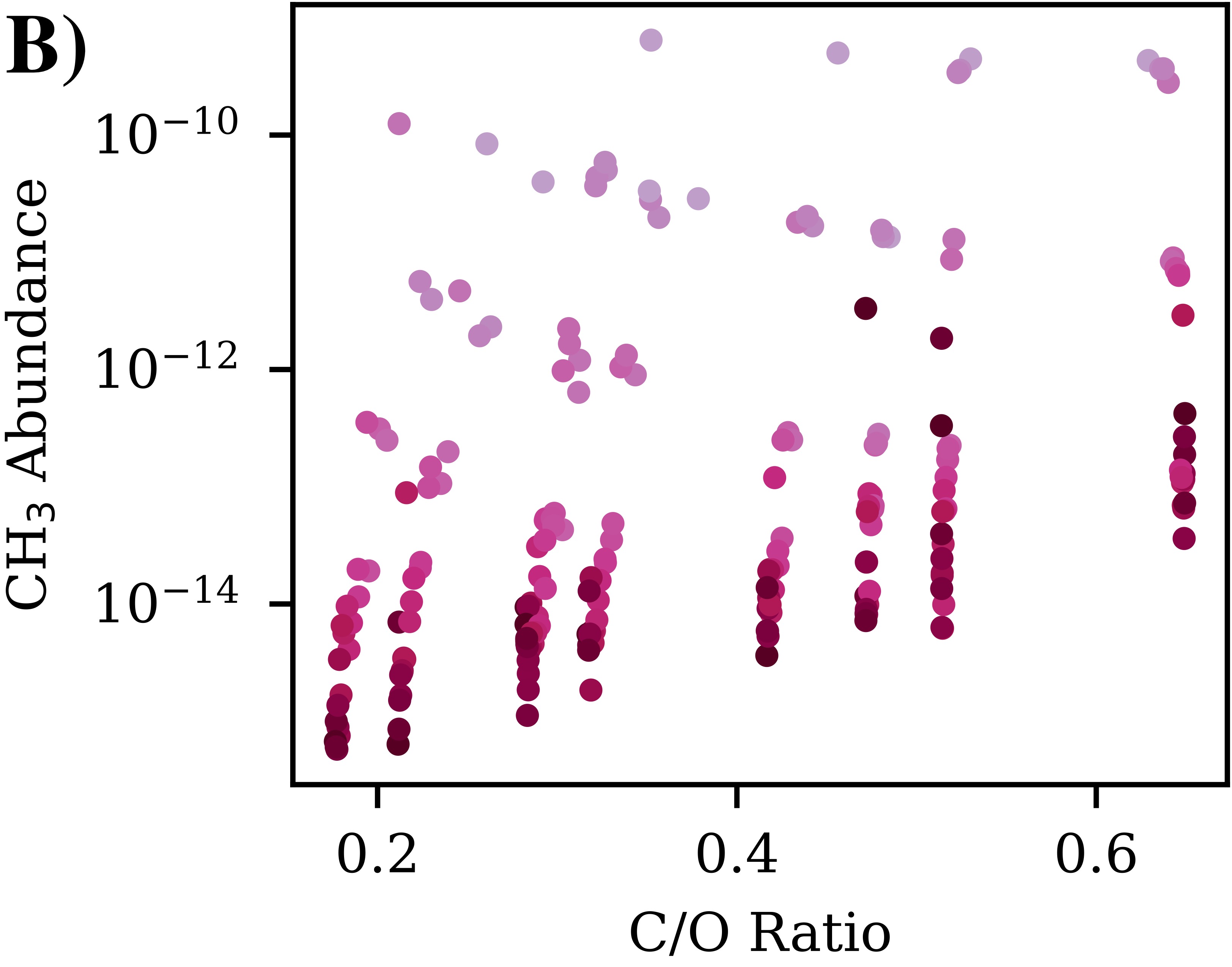}{0.75\columnwidth}{}}
\gridline{\fig{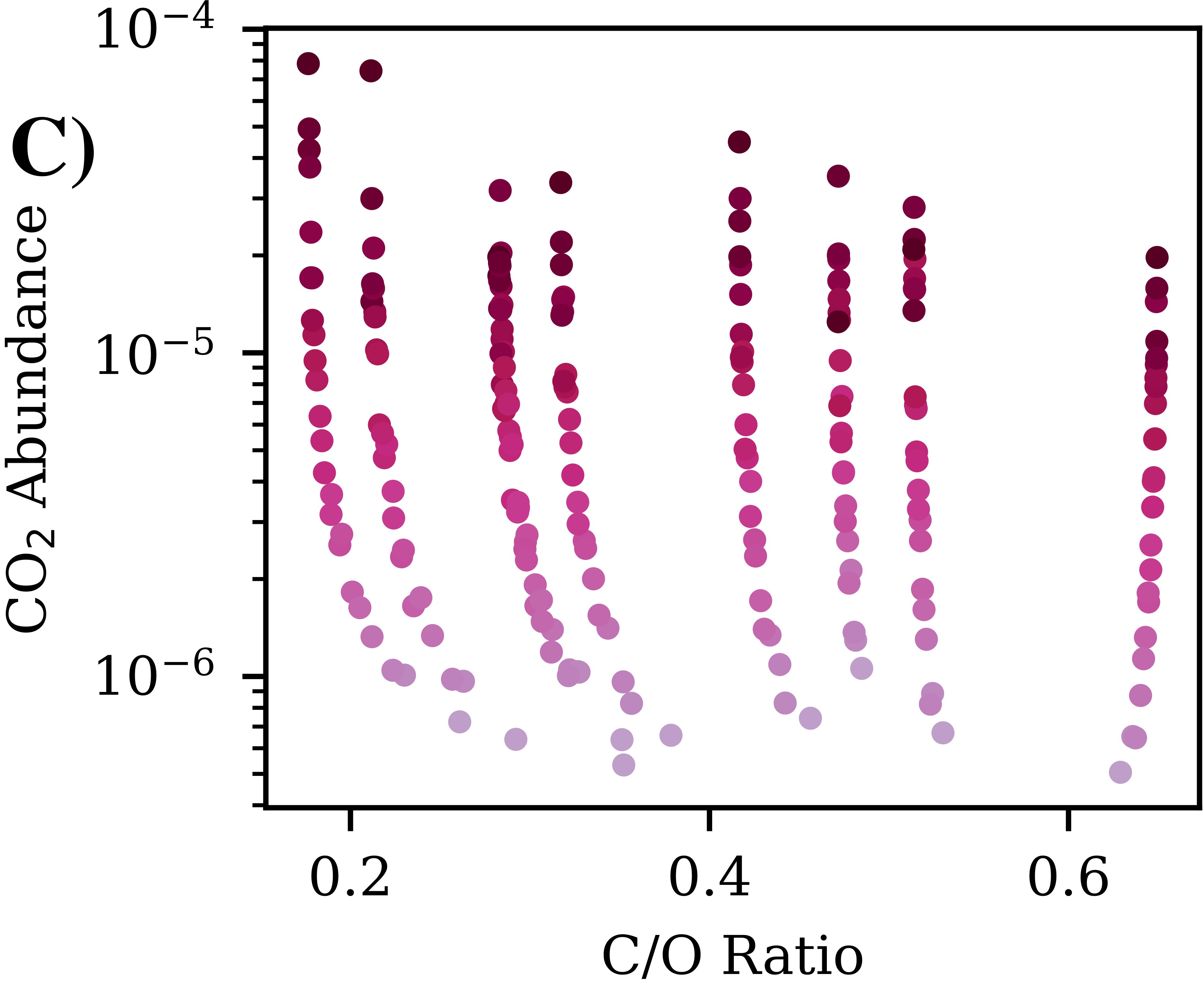}{0.75\columnwidth}{}
          \fig{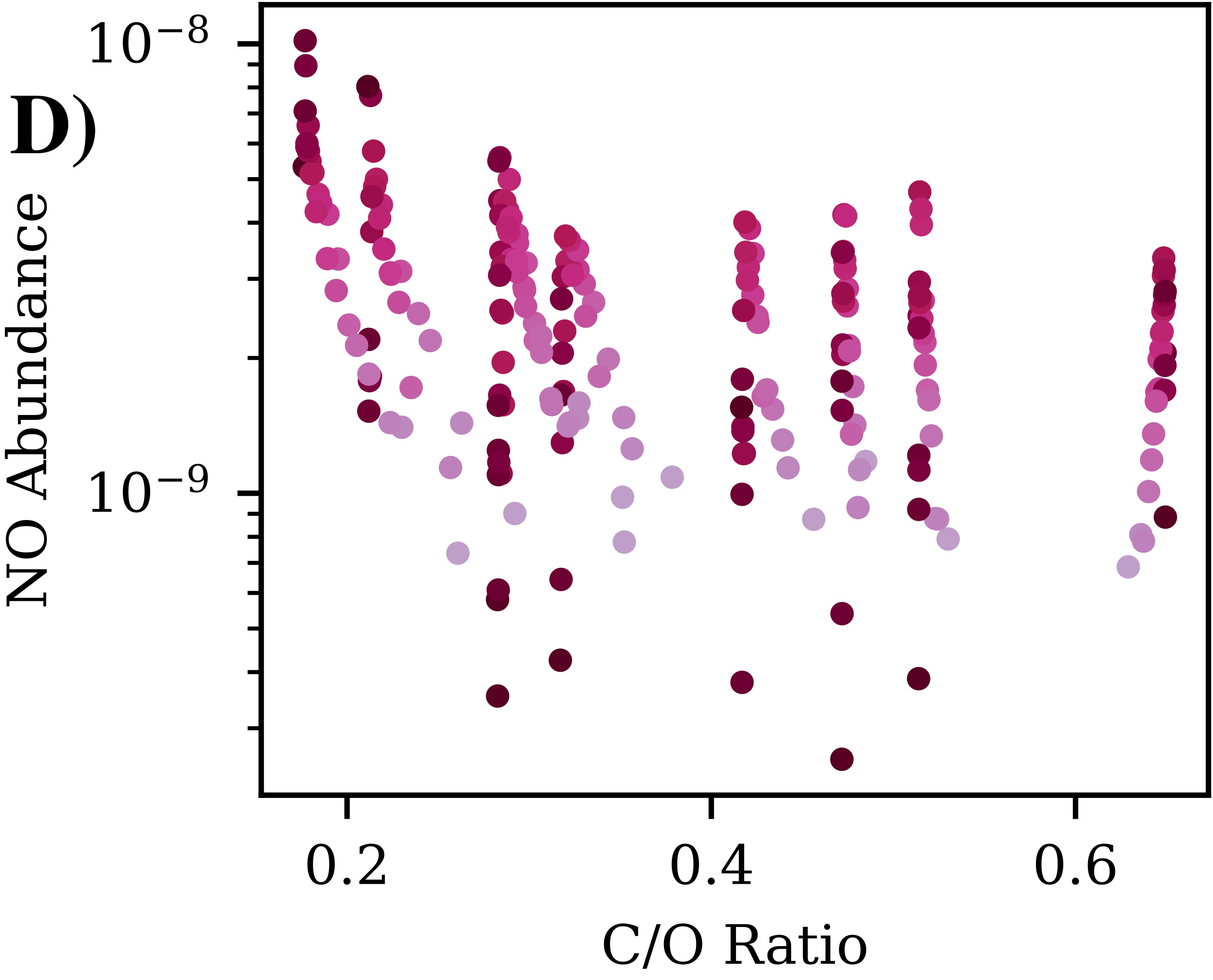}{0.75\columnwidth}{}}
\caption{Select mean upper atmosphere ($P<10^{-2}$ bar) fractional abundance versus C/O ratio scatter plots for the full suite of models considered here, i.e., for models with abundances discussed in \autoref{sec:comet_param}, radii of between 2.5 km and 30 km, and densities of 1 or 2 g cm$^{-3}$. Here we focus on four molecules, H$_2$O (A), CH$_3$ (B), CO$_2$ (C), and NO (D), with the first three acting as an example of one the main regimes found in our full analysis: decrease/increase/no-change in abundance with C/O ratio respectively, with the abundance trend showing an increase or decrease with cometary mass. {In the top left figure (A) we have labelled which row of \autoref{tab:comet_models} each abundance-C/O curve belongs to.} On the other hand, NO is representative of the reversal found for many nitrogen-bearing molecules. Note that the abundance-C/O and scatter regimes for each molecule with an upper atmosphere abundance $>10^{-16}$ are given in \autoref{tab:full_comp}, and the corresponding plots are available in the appendix.  \label{fig:full_comp_co}}
\end{figure*}
\begin{table*}
\resizebox{\textwidth}{!}{
\begin{tabular}{c|ccc}
 & \makecell[l]{High ($>10^{-8}$)} & \makecell[l]{Medium ($>10^{-12}$ and $<10^{-8}$)}  & \makecell[l]{Low ($>10^{-16}$ and $<10^{-12}$)} \\ \hline
\makecell[l]{Positive}   &\makecell[l]{{\bf CO}, C, HCN, }& \makecell[l]{CH, $^3$CH$_2$, CH$_4$, C$_2$H$_2$, C$_2$H$_4$, CH$_3$, \\ CN, {\bf HCO},} & \makecell[l]{$^1$CH$_2$, H$_2$CO, CH$_2$CO, CH$_3$OH, \\ C$_2$H, C$_2$H$_3$, H$_2$CN, HOCN, \\ HNCO, C$_2$N$_2$, {\it HCNH}, NCN,  } \\
&&&\\
\makecell[l]{Negative}     &\makecell[l]{{\bf H$_2$O}, {\bf OH}, {\it NO},}& \makecell[l]{{\bf O($^3$P)}, {\bf O$_2$}}& \makecell[l]{{\bf O($^1$PD)}, {\it HNO}, CO$_2$H } \\
&&&\\
\makecell[l]{No Trend}     &\makecell[l]{H$_2$, {\bf CO$_2$}, {\bf H}, He,\\ {\bf N$_2$}, N(4S),}&\makecell[l]{NH, NH$_2$, NH$_3$, }&\makecell[l]{N(2D), {\bf NNH}, N$_2$O, N$_2$H$_2$ }               
\end{tabular}
}
\caption{A table showing how, for all of our cometary compositions, the mean {\bf upper} atmosphere ($P<10^{-2}$ bar) abundance versus C/O ratio trend changes for molecules with upper atmosphere fractional abundances $>10^{-16}$. Here molecules are categorised using three different indicators. The first is if their abundance increases, decreases, or is unchanged as the atmosphere C/O ratio changes (rows). The second is set by their peak abundance, with a peak abundance of $>10^{-8}$ being labelled as `High', a peak abundance of $>10^{-16}$ and $<10^{-12}$ being labelled as `Low', and any peak abundances that fall between these two regimes being labelled as 'Medium'. And finally the third is if their abundance increases (bold) or decreases (regular) with cometary mass, with outliers shown in italics.}
\label{tab:full_comp}
\end{table*}


\subsection{Mean Abundances and Atmospheric C/O Ratios} \label{sec:atmospheric_co_metal}

In order to study the full ensemble of more than 400 cometary impact models considered here, we next calculated the `global' atmospheric C/O ratio for each model and the upper atmosphere ($P<10^{-2}$ bar) mean fractional abundance for each molecule in each model. Note that we focus on the upper atmosphere since it is this region that is potentially observable with transmission spectra (i.e., it is here that the opacity typically reaches unity - see \autoref{sec:observational_implications}).  We also limit ourselves to molecules with upper atmosphere abundances $>10^{-16}$ since lower abundance molecules are not only highly sensitive to slight perturbations in the chemistry (driven by small number statistics), but are also highly unlikely to be detectable. 
Abundance-C/O maps and trends for the every molecule with upper atmosphere abundances that fulfil this criterion are shown in the extended version of \autoref{fig:full_comp_co} and given in \autoref{tab:full_comp} respectively.  \\
It is important to note that our ensemble of models only samples a limited region of the C/O parameter space. For the nine differenet cometary compositions under consideration, there are eight different cometary ice C/O ratios since the planetesimal models differ only in nitrogen content. Furthermore, as the size of the impactor increases, the relative effect that it has on the atmospheric composition/density also grows, leading to the overall atmospheric C/O ratio slowly converging towards that of the incoming material. This can be seen in \autoref{fig:full_comp_co}, where we find eight distinct Abundance-C/O ratio curves. In a way, the steps along each curve are indicative of how ongoing bombardment by material with very similar compositions (or C/O ratios) might change the global chemistry of a hot Jupiter. \\

Here, as shown in \autoref{fig:full_comp_co}, we focus on four molecules which are exemplary of most features found for our ensemble of cometary impact models. Generally, these features are either an increase or decrease in abundance with increasing atmospheric C/O ratio and/or impactor mass (and hence atmospheric metallicity). But there are a number of molecules whose behaviour falls outside these bounds. For example the upper atmosphere abundance of CO$_2$ is essentially independent of the C/O ratio, only being affected by the increase in metallicity associated with cometary impacts. 
We also find that, as discussed in \autoref{sec:steady-state} and \autoref{sec:size}, the behaviour of nitrogen-bearing species follows its own trends. \\
We now look at these trends in more detail, starting with oxygen-rich molecules. As shown in \autoref{fig:full_comp_co}A for H$_2$O, we find that the abundance of oxygen-rich molecules increases with decreasing atmospheric C/O ratio (and increasing metallicity). This occurs because, at lower C/O ratios, the amount of free oxygen available rises, leading to reactions that form oxygen-rich molecules being preferred over the formation/presence of hydrocarbons. At the same time, regardless of cometary composition, we find an increase in the abundance of most oxygen-bearing molecules with impactor size. This effect is especially strong for water since, apart from one theoretical composition, water is the primary constituent of our cometary ices and hence is directly delivered, in large quantities, during an impact.
Briefly this occurs because cometary impacts change not only the C/O ratio of the atmosphere, but also the metallicity, especially the C/H, O/H and N/H number ratios. This is because these ratios are orders of magnitude higher in cometary ice than in an unpolluted, H$_2$ dominated, hot Jupiter atmosphere.\\
Metallicity-driven effects also explain the trend found for CO$_2$, as shown in \autoref{fig:full_comp_co}C. Here we find little to no change in upper atmosphere abundance with C/O ratio, but strong abundance dependence on the impactor mass, and hence the overall carbon/oxygen content of the atmosphere.  \\
Next, we have a number of molecules, typically hydrocarbons such as CH$_3$, as shown in \autoref{fig:full_comp_co}B, which increase in abundance with C/O ratio but decrease in abundance with impactor size. The former effect is simply the result of additional atmospheric carbon leading to more hydrocarbons forming, whilst the latter effect occurs because, regardless of composition, the cometary C/O ratio is always $<1$, hence more oxygen is always deposited than carbon. \\
Finally, we find that the upper atmosphere abundances of most nitrogen-bearing molecules such as NO, as shown in \autoref{fig:full_comp_co}D, behave rather differently to the above trends. At low impactor masses, we find either a simple increase or decrease in abundance with impactor size. However as this size grows, we eventually cross a threshold and the trend reverses. In order to understand why this occurs, it is important to note that, as seen in \autoref{fig:JWST_ICE_A_multi_comp}C/D, such reversals are limited to the upper atmosphere, vanishing when we calculate the global mean. Thus, the local nature of this reversal combined with the universality of this trend (i.e. it occurs even when comets do not introduce nitrogen to the atmosphere), suggests that the underlying cause is photochemistry.  Not only is the upper atmosphere very highly UV irradiated, but all of the carriers of nitrogen in our model are either photochemically active or directly form from molecules that result from the photodissociation of other abundant molecules, such as water. 
For example, at early times, the strong photolysis of incoming water leads to a strong initial switch from NH$_3$ towards NO and NO$_2$ in the upper atmosphere. This is then vertically mixed, resulting in the lower atmosphere enhancements seen in \autoref{fig:JWST_ICE_A_multi_comp}C/D.  However, as the amount of water in the upper atmosphere grows, so too does the strength of this initial photolysis (see \autoref{sec:non_linear_water}), leading to increasingly strong initial NH$_3$ depletion and NO/NO$_2$ enhancement. 
In turn, the compositional gradients for these molecules between the upper and lower atmospheres also grows, strengthening vertical mixing and hence further enhancing the deep NO and NO$_2$ enrichment. 
The relatively high mass of NO and NO$_2$ molecules then bakes this deep enrichment in. Since both NO and NO$_2$ are heavy, they tend to stay in (or sink into) the lower atmosphere, even when the compositional gradients which drove the initial mixing/diffusion have disappeared. On the other hand NH$_3$ is much lighter, and hence is more freely mixed between the upper and lower atmospheres as the simulation progresses. 
Thus the switch in NH$_3$ abundance in the upper atmosphere can also be linked to the non-linear change in water abundance with impactor mass. For the most massive impactors, the water content of the atmosphere is actually reduced, hence more NH$_3$, which was sequestrated in the mid/lower atmosphere at earlier times, can be mixed back into the upper atmosphere and survive, leading to the reversal/enhancement seen in \autoref{fig:JWST_ICE_A_multi_comp}C. \\
Such a scenario also explains the difference in reversal threshold impactor masses for different compositions (\autoref{fig:full_comp_co}D). As the cometary composition changes so too does the water (oxygen) and nitrogen content of the upper atmosphere, impacting reaction (production) and photolysis rates and hence the point at which we switch between enhancement and depletion. This conclusion is reinforced by our analysis (not shown) of the change in abundance of nitrogen-bearing molecules with N/O ratio. At `low' N/O ratios. oxygen bearing nitrogen species are preferred, whereas at `high' N/O ratio, hydronitrogens become increasingly abundant. This is true in both the upper atmospheres as well as when taking a `global' average. \\
Note that a similar effect, vis-a-vis photolysis and heavier molecules sinking out of the upper atmosphere also explains the abundance trends of many oxygen-bearing organic molecules, such as CH$_3$OH, which show depletion/enhancement with impactor mass when we look at the upper or `global' atmospheric averages respectively.

\begin{figure*}
\gridline{\fig{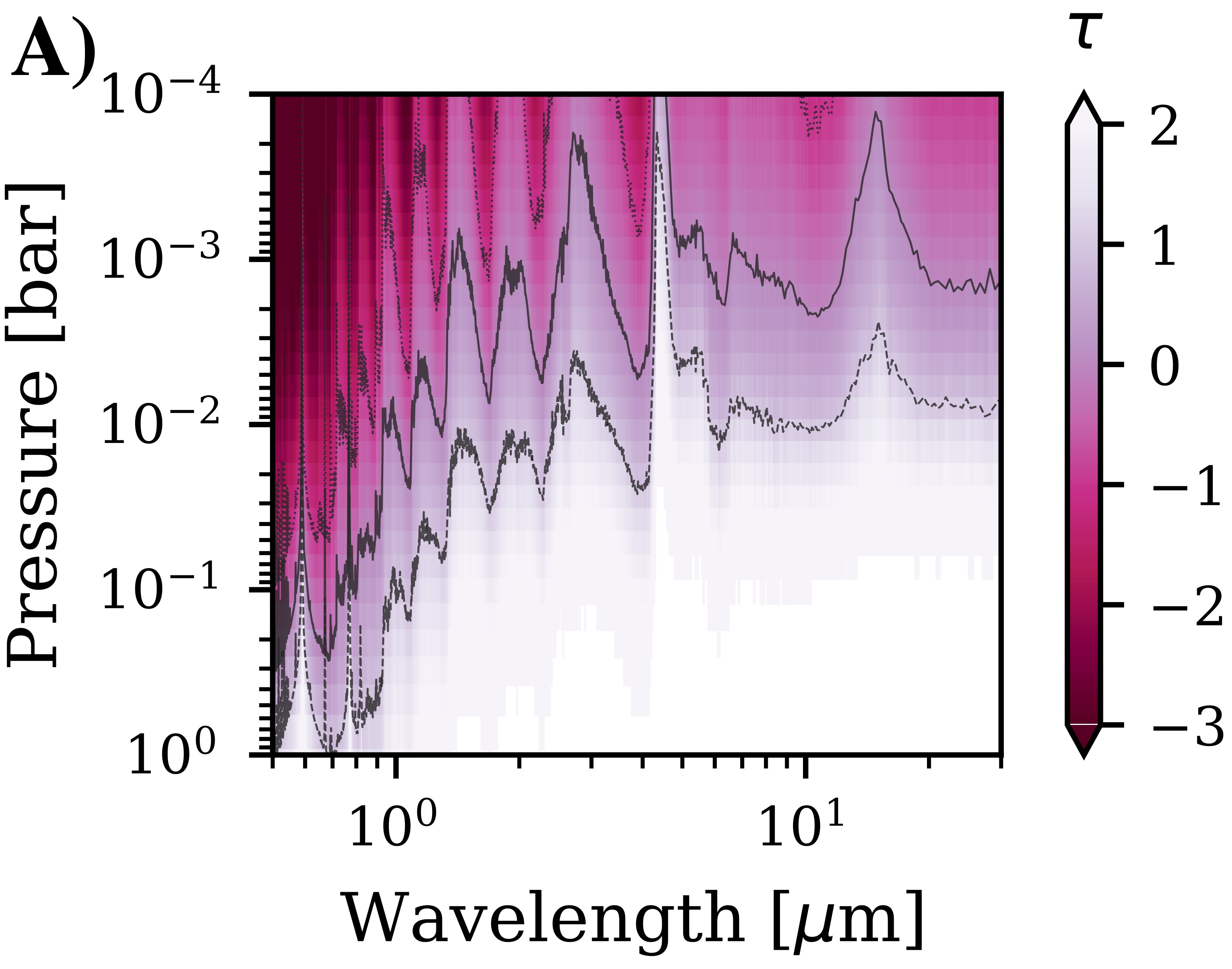}{0.45\textwidth}{} 
          \fig{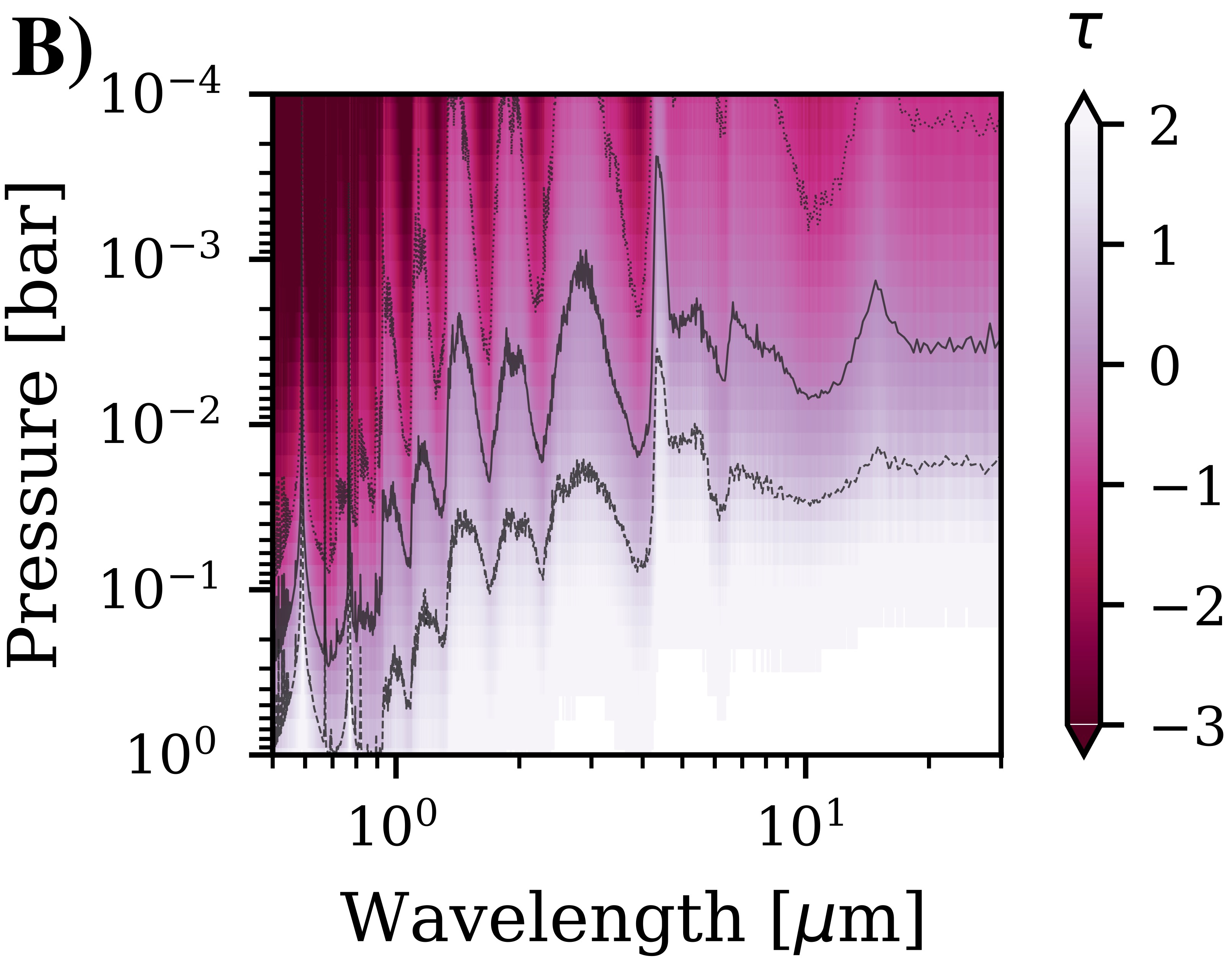}{0.45\textwidth}{}} 
\gridline{\fig{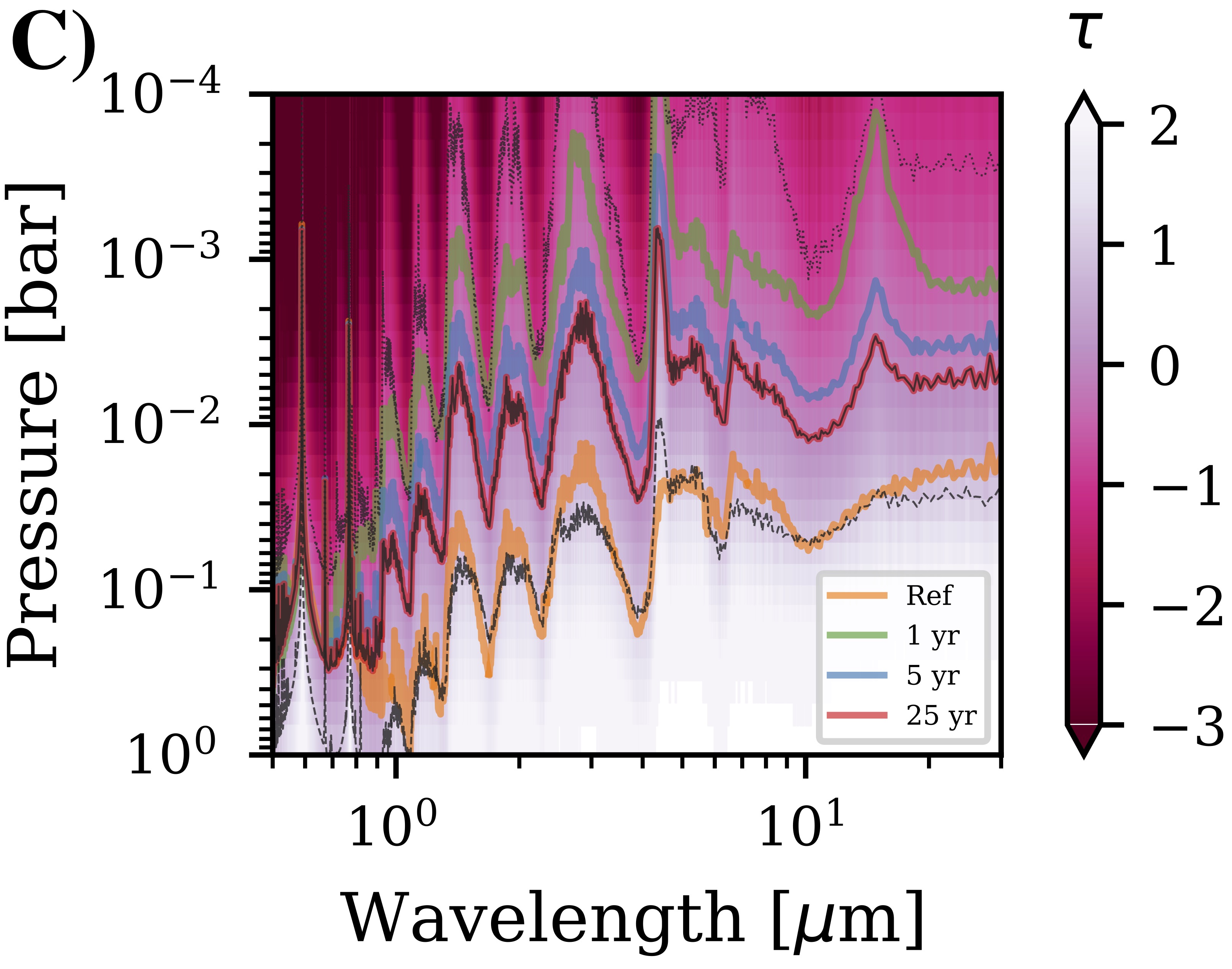}{0.45\textwidth}{} 
          \fig{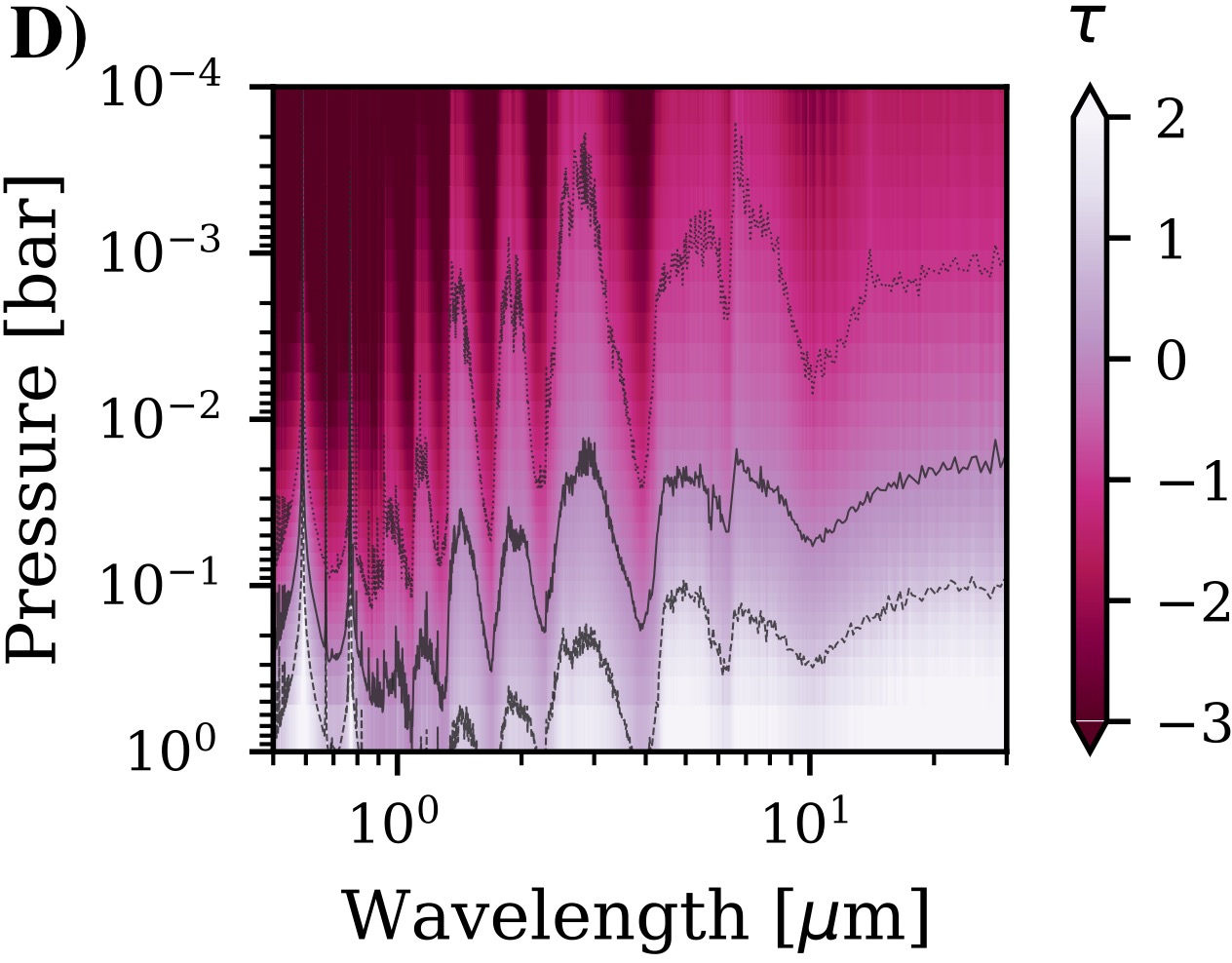}{0.45\textwidth}{}} 
\caption{Maps of the wavelength dependent optical depth (on a log scale), with optical depths of 0.1, 1.0 and 10 indicated by dotted/solid/dashed line respectively. We compare the optical depth profiles for an R$=$10 km, $\rho=$2 g cm$^{-3}$, JWST Ice ver. A cometary composition impact at three different points in time, 1 year (A), 5 years (C), and 25 years (C - steady-state) after impact, with our unperturbed HD209458b reference model (D). {To emphasise how the opacity of the atmosphere varies time we also include, on our steady-state opacity map, the 1 year (green), 5 year (blue), 25 year (red) and reference (orange) $\tau=1$ surfaces as light shading.} Note the significant effect that the addition of cometary material has had on the optical depth, in particular how an optical depth of one is reached at lower pressures, averaging $\sim 10^{-2}$ bar at steady-state vs $\sim 10^{-1}$ bar in the unperturbed model, helping to explain the difference in the transmission spectra seen in \autoref{fig:JWST_ICE_A_10KM_2g_transmission}. \label{fig:JWST_ICE_A_10KM_2g_opacity}}
\end{figure*}
\begin{figure}
 \gridline{\fig{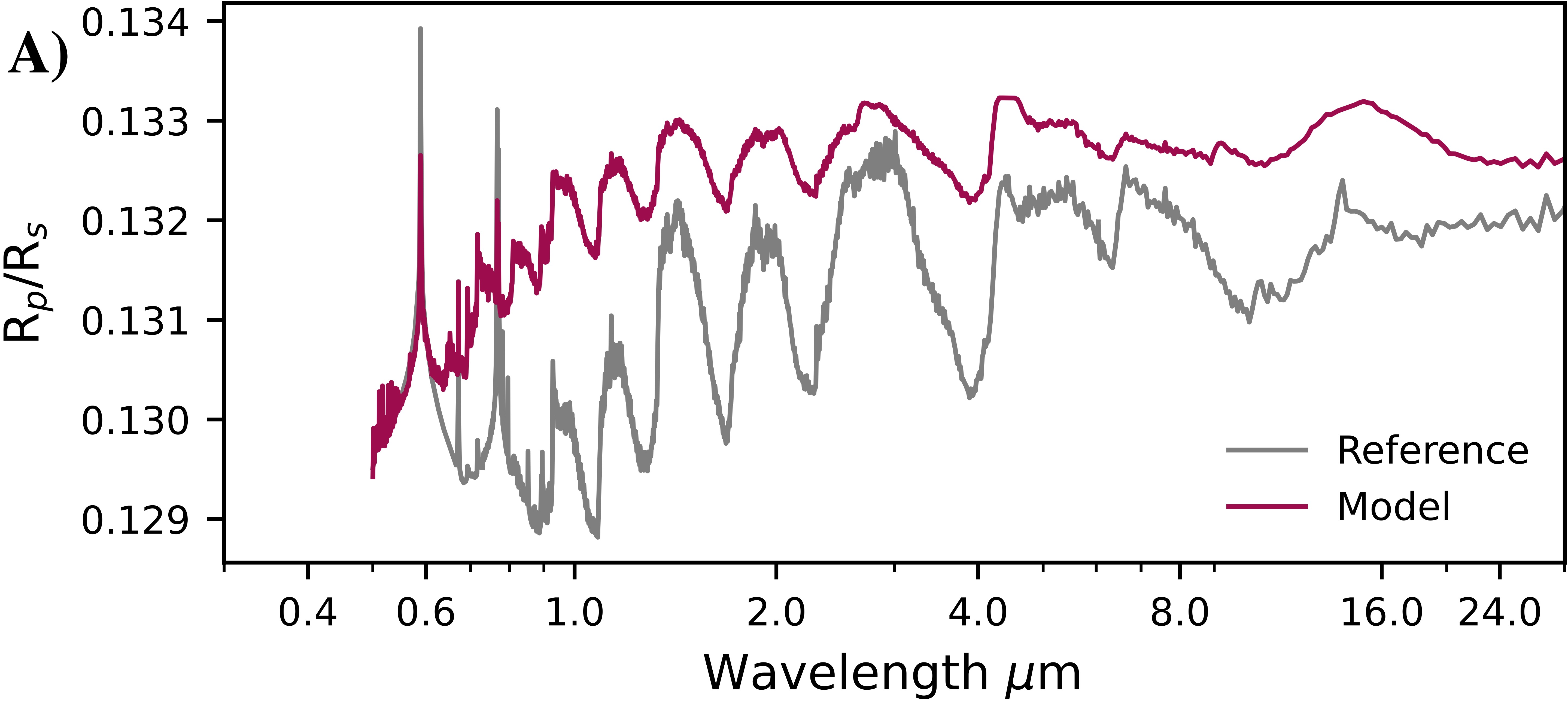}{0.45\textwidth}{}} 
\gridline{\fig{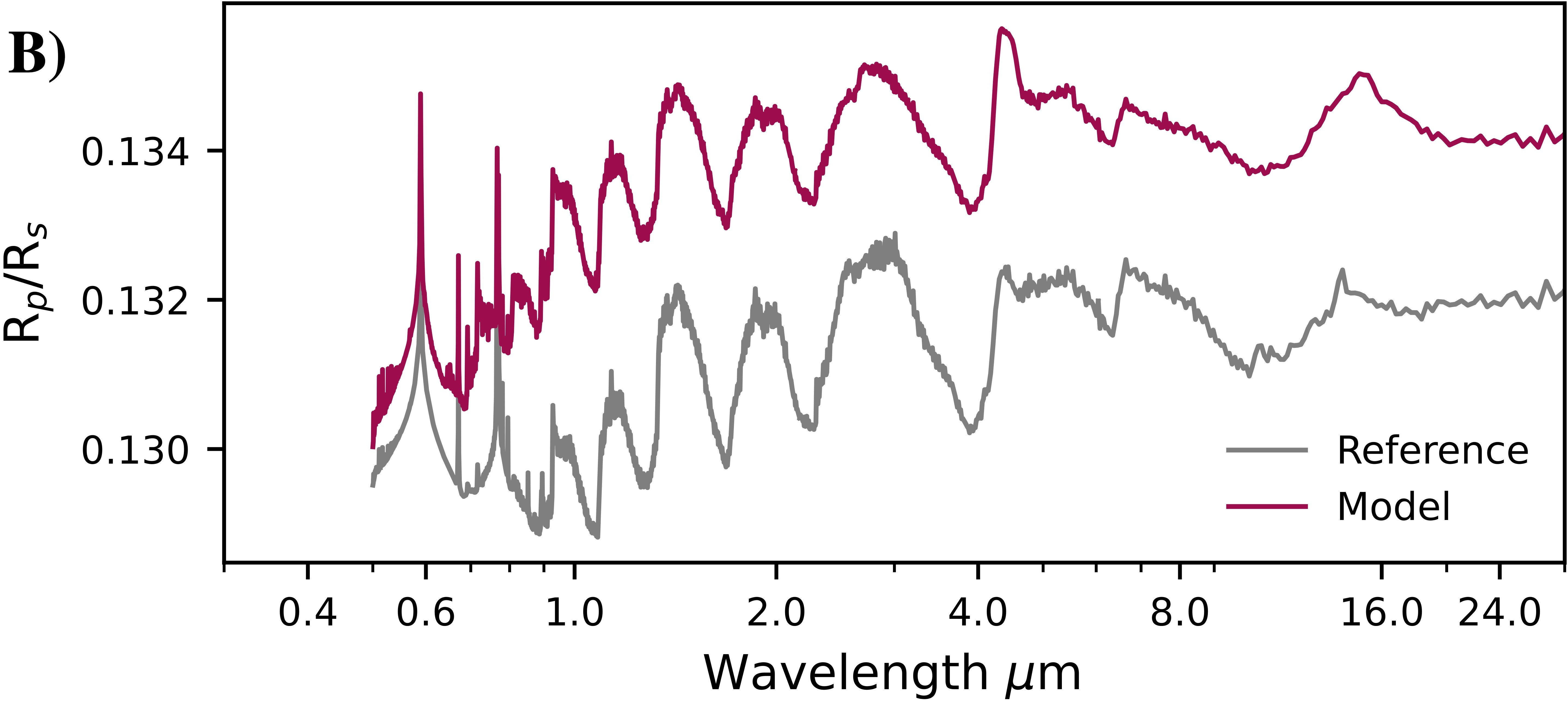}{0.45\textwidth}{}} 
\gridline{\fig{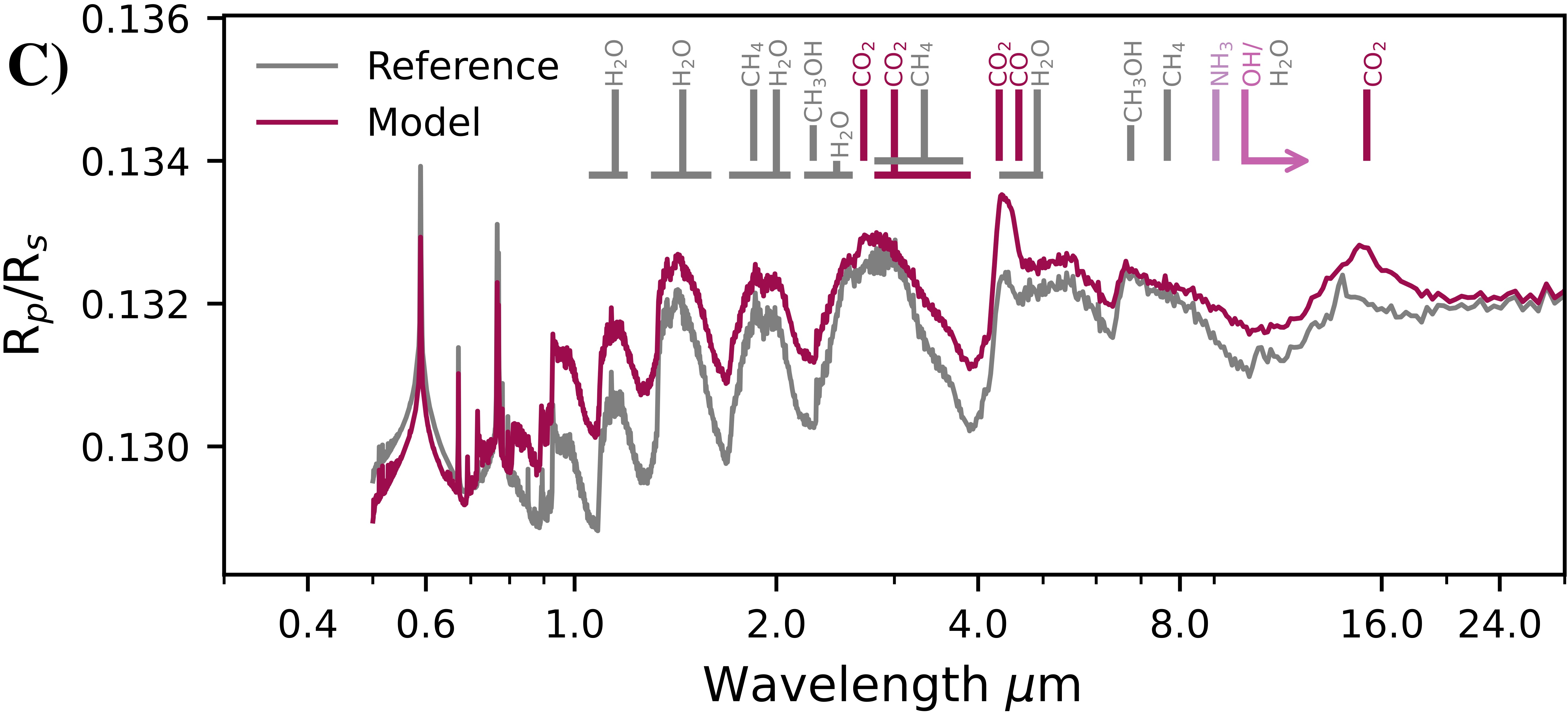}{0.45\textwidth}{}} 
\caption{Example transmission spectra R$=$10 km, $\rho=$2 g cm$^{-3}$, JWST Ice ver. A cometary composition impact at three different points in time, 1 year (A), 5 years (B), and 25 years (C - steady-state) after impact, selected to illustrate the potentially observable effect that a cometary impact has on the planetary atmosphere both locally on short time-scales as well globally after a period of sustained/significant bombardment. We also include, as a comparison, the spectrum from our unperturbed HD209458b model. At steady-state, between 1 $\mu$m and 15 $\mu$m, we have labelled a number of features of interest, with stronger than reference features shown in red and weaker than reference features shown in grey - with the saturation of the colour indicating our confidence in the identified feature. \label{fig:JWST_ICE_A_10KM_2g_transmission}}
\end{figure}
\subsection{Potential Observability} \label{sec:observational_implications}

To understand how the effects of cometary impacts might be detected, it is important to understand how the aforementioned vertical differences in atmospheric chemistry and temperature impact observable features. To that end, we now focus on a single cometary impact, with R$=$10 km, $\rho=$ 2 g cm$^{-3}$, and JWST Ice ver. A composition, exploring the time-dependent changes to synthetic opacity maps (optical depth - \autoref{fig:JWST_ICE_A_10KM_2g_opacity}) and transmission spectra (\autoref{fig:JWST_ICE_A_10KM_2g_transmission}) and comparing them to an unperturbed HD209458b atmospheric model. Due to the 1D nature of our models, which mean that cometary material remains in the model instead of being mixed with the unimpacted surroundings (a process that occurs over tens of years, as seen for SL9), we treat different points in time as being representative of different scenarios. \\
Early points in time, e.g. 1 year and 5 years after impact, are considered to be an {optimistic example} of the highly localised changes that {might} occur shortly after impact, before any dissipation/mixing occurs. As such, observations similar to what we discuss for these points in time represent a best case scenario in which the cometary impact is recent, occurs close to or on the terminator, and alters the {terminator region} chemistry enough to be observable. This is similar to the short-lived changes observed in the 2.3 $\mu$m methane feature observed after the impact of comet SL9 with Jupiter, which \citet{2016Icar..264....1F} linked to local (particulate) material delivery and a resulting change in the opacity and albedo of Jupiter's upper atmosphere. {Such changes effectively covered the atmosphere at the 45$^{\circ}$ latitude of Jupiter \citep{2016Icar..264....1F} and we assume a similar scenario occurs for the terminator region of our impacts models. Consequently our 1 year and 5 year impact models likely overestimate the effect that a single impact has on the terminator region. Yet we still discuss them here because the effects observed have important implications for the effect of cometary material on planetary atmospheres, reinforcing the need to pair observations with a robust atmospheric model. }   \\
On the other hand, we consider the steady state (25 years after impact for this model) to be representative of an atmosphere that has undergone continuous and/or ongoing bombardment by comets with similar compositions, leading to strong changes in the global composition/ chemistry of the atmosphere. Such a scenario might occur, for example, during and after the migration of a hot Jupiter from its formation location to its current tidally-locked orbit. \\

We start, in \autoref{fig:JWST_ICE_A_10KM_2g_opacity}, by exploring how our cometary impact changes the atmospheres optical depth ($\tau$). Once the model has settled down slightly from the initial perturbation, i.e. after the first few months of simulation time, we find a prolonged period in which the atmospheres opacity is significantly (\autoref{fig:JWST_ICE_A_10KM_2g_opacity}A) enhanced with respect to our unperturbed model, with $\tau=1$ being reached at an average pressure of $\sim 10^{-3}$ bar versus $\sim 10^{-1}$ bar in the unperturbed model. 
The underlying cause of this spike in opacity appears to be the strong H$_2$O (and to a lesser extent CO and CO$_2$) enrichment that occurs during the cometary impact. This material acts as a strong opacity source, absorbing incoming radiation and causing local heating of the upper atmosphere, an effect which is also slightly enhanced by the initial heating associated with the cometary impact. In turn, the hotter upper atmosphere strengthens H$_2$-H$_2$ and H$_2$-He collision-induced-absorption (CIA), which is highly temperature dependent \citep{BORYSOW2001235}, shaping the increase in optical depth seen in \autoref{fig:JWST_ICE_A_10KM_2g_opacity}A/B/C for almost all wavelengths except those associated with K/Na absorption. \\
Over time, as the initial upwelling of water settles or is photo-dissociated, the opacity of the upper atmosphere drops, leading to cooling and hence a weakening of CIA. This can be seen in \autoref{fig:JWST_ICE_A_10KM_2g_opacity}B as an increase in the pressure at which the atmosphere becomes optically thick ($\tau=1$). \\
This process continues as the atmosphere evolves towards steady-state (as shown in \autoref{fig:JWST_ICE_A_10KM_2g_opacity}C), with the depth at which $\tau=1$ reaching an average of $\sim 10^{-2}$ bar. Yet this is still much shallower than the average $\tau=1$ pressure, $\sim 10^{-1}$ bar, in an unperturbed HD209458b-like atmosphere, reflective of the enhanced water content (\autoref{sec:steady-state}) and upper atmosphere temperature (\autoref{sec:PT}) driven by the cometary impact. \\
Note, however, that as previously alluded to, this enhancement in opacity is not uniform with different components of the atmosphere affecting the opacity at different wavelengths. For example between 4 and 5 $\mu$m all of our atmospheric models exhibit an almost order of magnitude spike in optical thickness, a spike which is absent for HD209458b. This is caused by CO and CO$_2$, which have strong absorptions at $4.6\mu$m and $4.2\rightarrow4.4\mu$m respectively, and which are both significantly enhanced (\autoref{sec:steady-state}) in abundance when compared with our unperturbed atmosphere. \\

To explore these differences in more detail, we now turn to synthetic transmission spectra (\autoref{fig:JWST_ICE_A_10KM_2g_transmission}). \\
One year after impact, we find a transmission spectrum spectrum (\autoref{fig:JWST_ICE_A_10KM_2g_transmission}A) that remains significantly impacted by the still settling cometary material.  In particular, we find that {whilst} the apparent planetary radius, $R_{P}/R_{S}$, is enhanced at almost every wavelength, {the strength of most (other than the $4\rightarrow5\mu$m CO/CO$_2$ or the $16\mu$m CO$_2$ features) absorption features is reduced, with respect to our unperturbed HD209458b atmosphere. This combination suggests that the planet itself is not larger/inflated, but that instead we are probing lower pressure regions of the atmosphere, where the densities of most molecules causing the observed absorption features is reduced. This is exactly what the opacity maps discussed above (\autoref{fig:JWST_ICE_A_10KM_2g_opacity}A/D) reveal: the $\tau=1$ region, which our synthetic transmission spectrum probes, occurs at a much lower pressure for our impacted model than for our unperturbed model. Furthermore, the shape of this change to the transmission spectrum (and opacity maps) is highly reminiscent of the high-temperature H$_2$-H$_2$ CIA absorption curve seen in figure 1 of \citet{BORYSOW2001235}. Thus the absorption features seen in \autoref{fig:JWST_ICE_A_10KM_2g_transmission}A are a combination feature of CIA and H$_2$O opacities, and the enhanced CIA opacity means that less water molecules are needed before the atmosphere becomes optically thick, hence reducing the relative strength of the water absorption feature. This conclusion is reinforced by the strengthening of the CO$_2$/CO feature, which occurs because the relative enhancement of these molecules is so boosted} by the impacting comet (peaking at a factor of $10^{6}$ for CO$_2$) that any reduction {in probed} atmospheric density is more than balanced out by the increase in fractional abundance. \\

Five years after impact, although the initial upper atmosphere H$_2$O enhancement has started to significantly settle/photodissociate, and the atmosphere has become optically thinner (\autoref{fig:JWST_ICE_A_10KM_2g_opacity}B), the apparent radius of the planet has continued to increase (\autoref{fig:JWST_ICE_A_10KM_2g_transmission}B){, and so too has the strength of the absorption features, particularly the H$_2$O, CO and CO$_2$ features}. {This occurs because the very outer atmosphere has cooled slightly (as water settles), weakening the very low pressure H$_2$-H$_2$ CIA opacity, and thus enabling us to probe higher pressure/density/temperature regions of the atmosphere, leading to enhancements in} both molecular absorption and CIA effects.  This is reflected in both the peak to trough feature strength, which has a maximum around twice that found one year after impact, as well as the `continuum' level, which tracks the strength of the deeper/hotter H$_2$-H$_2$ CIA rather than just being driven by the altitude at which the atmosphere becomes optically thick. The subtlety of this effect reinforces the value of exploring synthetic transmission spectra over just looking at abundances and opacities. It is the best way to understand how vertical differences in composition and temperature impacts observations. \\
Note that the above increase in apparent radius, and hence the asymmetry in observed planetary radii between eastern and western terminators, is one of the likeliest ways that a very recent, near terminator, impact might be identified. And such a scenario is not unlikely, as discussed in \citet{1982stjp.conf..277S} and \citet{ZAHNLE2001111}, the cratering (impact) rate at the apex of motion (i.e. the eastern terminator) can reach 2.6 times the planetary average, whilst also dropping by a factor of 15 on the trailing edge (i.e the western terminator).   \\

Whilst the aftermath of a single impact is highly unlikely to be captured, ongoing, or a period of significant, bombardment by comets with similar compositions might lead to observable changes in the global atmospheric composition/chemistry. We explore such a scenario using a steady-state atmospheric model whose synthetic transmission spectra in shown in \autoref{fig:JWST_ICE_A_10KM_2g_transmission}C. This spectra is typical of that found throughout our ensemble of steady-state atmospheric models, all of which increase the water (and hence oxygen) content of the atmosphere. Similar to what was found at earlier times, the influx of high opacity material, such as H$_2$O, CO and CO$_2$ has, via atmospheric heating and CIA effects, resulted in a general enhancement in the continuum level with respect to our unperturbed atmosphere, whilst also weakening many atmospheric features, such as H$_2$O, K and Na, due to our observations probing a lower density region of the atmosphere (\autoref{fig:JWST_ICE_A_10KM_2g_opacity}C/D). \\
Note that the above enhancement of the continuum level is highly correlated with the atmospheric water abundance and hence impactor mass, leading to some of our higher cometary mass models showing significant enhancements in the apparent radius of the planet, at least when integrating over the wavelength range considered here. This is important because some observatories, including JWST, observe exclusively in the near/mid infrared and as such an exaggerated radius might be reported \citep{2016ApJ...832...41M}. There are two solutions to this problem. The first is to take additional observations with either a UV or optical telescope, both of which are less sensitive to these water and CIA driven effects. Whilst the second is to pair these observations with a robust atmospheric model that includes a complete treatment of atmospheric opacity sources. This works because a number of atmospheric features help to break the degeneracy between the observed `continuum' level and the planetary radius.  \\
Examples of these features have been labelled on \autoref{fig:JWST_ICE_A_10KM_2g_transmission}C and include: 
\begin{itemize}
    \itemsep0em 
    \item enhanced CO$_2$ absorption at 2.6$\mu$m $2.8\rightarrow3.9\mu$m, $4.2\rightarrow4.4\mu$m, and $15.1 \rightarrow 15.3\mu$m,
    \item enhanced CO absorption at $4.6\rightarrow4.7\mu$m,
    \item possible weak NH$_3$ absorption at $9.0\rightarrow9.1\mu$m,
    \item possible OH absorption at $>10\mu$m,
    \item reduction in CH$_4$ absorption at 1.9$\mu$m, $2.8\rightarrow3.8\mu$m, and $7.7\mu$m
    \item slightly suppressed H$_2$O absorption effects, due to probing lower density regions of the atmosphere, at $1.3\rightarrow1.6\mu m$, $1.7\rightarrow2.1\mu m$, and $\sim3\mu m$ etc..  
\end{itemize}
Many of these features fall into the region at which JWST is expected to be most sensitive, which is the fixed slit mode of NIRSPEC which has $R\rightarrow2600$ between 0.97 and 5.27$\mu$m \citep{2022A&A...661A..80J}. The mid-infrared CO$_2$ feature at $\sim15.2\mu$m is also potentially detectable with MIRI. \\
{Further, by comparing these features with potential efractory/metallic elements, such as K (0.77$\mu$m) and Na (0.58$\mu$m), which should be unaffected by cometary impacts, it is possible that we might be able to identify objects in which the oxygen, or carbon, metallicity has been enhanced by cometary material delivery. Of course, this would require us to know/measure the abundance of atmospheric K/Na in the first place, but a even a low estimated refractory to volatile ratio might be a good indicator that an object is worthy of further study via both observations and atmospheric modelling. \\ }
Such features, if identified and attributed to changes in atmospheric C/O ratio and metallicity, provide the most-likely pathway to identifying hot Jupiters that have either undergone significant cometary bombardment in their past, or are still being bombarded to this day. \\
{Note that we have focussed the above discussion on spaced based observations, such as those using JWST, which are are currently the best way to constrain/observe the presence of water on distant planets. However future ground based telescopes, such as ELT, might/will be able to break the veil of telluric contamination, and constrain the atmospheric C/O ratio and metallicity via high spectral resolution observations paired with robust atmospheric modelling, such as that available with ATMO. }

\subsection{Limitations/Caveats} \label{sec:limitations}
We reiterate that the above discussion of observations represents a best case scenario, either capturing a near terminator impact that very recently occurred or observing a hot Jupiter atmosphere which has been significantly polluted by massive or ongoing bombardment. Consequently, rather than treated as examples of cometary impacts that observers should go out and hunt for, they should be taken as optimistic examples of how material delivery by cometary impacts might reshape hot Jupiter atmospheres and break the link between current atmospheric composition and formation location.\\

Also, all of the results presented here are for {single impacts and} 1D atmospheric models. This adds some inherent uncertainty to some of our results and their applicability. {To alleviate these concerns, in addition to the models discussed throughout this work, we ran two additional models in which cometary material delivery was spread over ten smaller impacts. For our 5 km-equivalent model we found that the overall atmospheric abundances were very similar, just with a slight enhancement in CH$_4$ due to its continuous delivery. This effect became more pronounced as we move to the 22.5 km-equivalent model, where we find that continuous material delivery has resulted in an approximately seven times increase in overall CH$_4$ and NH$_3$ abundance, as well as a 15\% increase in atmospheric H$_2$O. The latter change suggests that the enhanced water photolysis discussed in \autoref{sec:non_linear_water} is indeed likely to be muted in a real hot Jupiter atmosphere, since horizontal mixing will have a similar effect to multiple weaker impacts, spreading the water out horizontally instead of temporally. Thus, combining both multiple smaller impacts (the continuous/ongoing bombardment case) with 3D mixing means that the enhanced H$2$O dissociation seen here is likely to be observed in only the most water enriched atmosphere. To truly understand these effects, in the future, we must turn to 3D atmospheric models, models which would also include vertical mixing, which may have the opposite effect to horizontal/temporal spreading, delivering material from the deep atmosphere to the very upper atmosphere where it photodissociates. \\ }
Note that a similar story might also impact the reversal in the abundance of nitrogen-bearing molecules in the upper atmosphere, although the fact that said reversal occurs at smaller cometary sizes, in our multi-impact model, and appears to be linked with molecular masses and vertical mixing mean that we are much more confident in the robustness of this feature.

\section{Concluding Remarks} \label{sec:discuss}

In this work we have implemented a simple cometary impact model, including ablation at low pressures and break-up deeper within the atmosphere, and coupled it with the `radiative-convective' atmospheric disequilibrium chemistry model, ATMO. This coupled model was then used to simulate over 400 cometary impacts with the day-side of the hot Jupiter HD209458b in order to investigate how the delivery of water-rich cometary material might affect the atmospheric structure, composition, and chemistry of such objects. \\
Due to the 1D nature of our models, we consider each impact model to be representative of two different scenarios: at early times the model is treated as being representative of the highly-localised changes caused by a single massive impact, whereas at later times, after mixing would be expected to dilute the changes associated with a single cometary impact in a 3D atmosphere, we consider our models to be representative of the steady-state reached after significant, global, bombardment by comets with similar compositions. \\
Using these models, we have shown that individual cometary impacts can drive significant, if short-lived, changes in local atmospheric chemistry, and that over longer periods of time, ongoing/continuous bombardment by comets might change the global atmospheric chemistry of hot Jupiters. The latter effect is of particular interest because many hot Jupiter formation location theories assume that the current C/O ratio of the atmosphere is reflective of the formation location, and hence assume that formation history can be simply backtraced from observations. Cometary material delivery can, and likely will, change this. Briefly, this occurs because comets are typically oxygen rich (i.e have low C/O ratios) and hydrogen poor (with O/H and C/H number ratios of order unity), and as such then can change both the C/O ratio and metallicity of the atmosphere, breaking the link between formation location and current day observations. And such a scenario is not unlikely because protoplanetary discs are not quiescent and the migration of a hot Jupiter is likely to be highly disruptive, potentially leading to cometary impact rates high enough to produce global atmospheric composition changes on the order of those found in our steady-state models. For example, modelling of the migration of planets within our own solar system suggests that the migration of Neptune through the proto-Kuiper-belt led to Jupiter `accreting' around 0.1 Earth masses of material, enough to significantly alter the chemistry of the mid/upper atmosphere (with P $<$ 10 bar) even after most of the incoming material has settled. In comparison, our smallest impact model changed the mass of our 1D column by only 1\%, yet led to significant, potentially observable, changes in synthetic transmission spectra and opacity maps. Consequently, {if, for example, JWST} reveals a wealth of planets with C/O ratios and/or atmospheric metallicities significantly different from those predicted by planetary formation or migration models, we propose that one mechanism by which we might understand such objects is the influx of material delivered by cometary bombardment. {And there are already signs that this might be the case, not only has JWST revealed a number of planets with super-solar metallicities, such as WASP-39b, which exhibits a strong SO$_2$ feature that requires 10x solar metallicity to reproduce \citep{wasp39bso2}, we also have increasing evidence that young planets, such as AF Lep b, have enriched metallicities, which could be due to bombardment \citep{2023AJ....166..198Z}.}  \\

The next step to understanding how cometary material delivery might break the link between planet formation location and atmospheric C/O ratio is to move beyond 1D atmospheric models, exploring how horizontal transport/mixing might affect our results. For example, the strong zonal jet that typically forms on tidally locked exoplanets might lead to significant day/night mixing, moving material from the irradiated day-side to the cold night-side where photolysis is suppressed. The problem is that the 3D GCMs which include the disequilibrium chemistry required to fully model cometary impacts (and hot Jupiter atmospheric chemistry more generally), are still in development or being validated. 2D atmospheric models represent a currently achievable compromise and come in two flavours: a pseudo-2D model in which we track a 1D atmospheric column as it is advected around the atmosphere by the equatorial jet, something which is achievable with very slight modifications to our current model, and a true-2D model which solves for disequilibrium chemistry and pressure-temperature profiles on the equatorial plane, and will require significant changes to be made to the chemistry solver implemented in ATMO.\\
Combining the above approaches with the wealth of anticipated JWST data giving unique insight into the composition of exoplanetary atmospheres, we foresee a bright future for understanding and investigating the potential effects of cometary impacts and material delivery on the chemistry and structure of hot Jupiter atmospheres.

\subsection{Appendix}
{Extended figures showing the abundance profiles and and C/O ratio variations of all potentially observable molecules.}

\begin{figure*}
\gridline{\fig{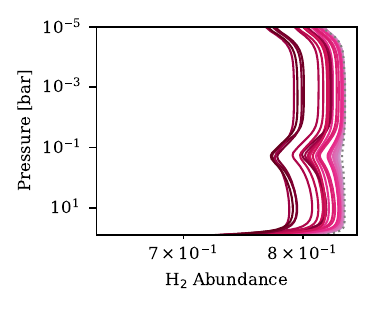}{0.5\columnwidth}{}
\fig{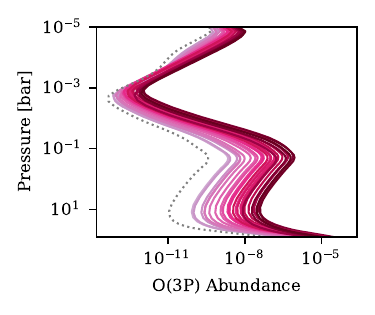}{0.5\columnwidth}{}
\fig{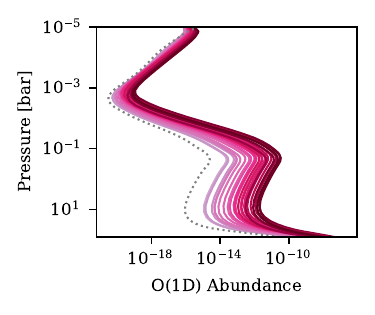}{0.5\columnwidth}{}}
\gridline{\fig{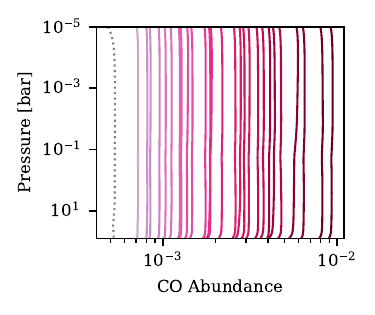}{0.5\columnwidth}{}
\fig{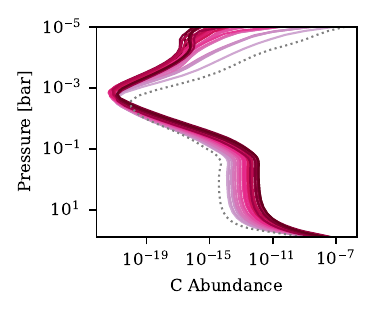}{0.5\columnwidth}{}
\fig{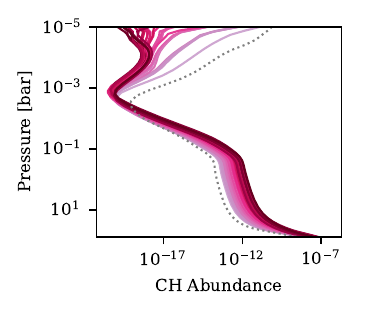}{0.5\columnwidth}{}}
\gridline{\fig{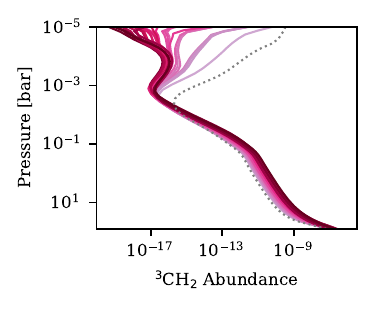}{0.5\columnwidth}{}
\fig{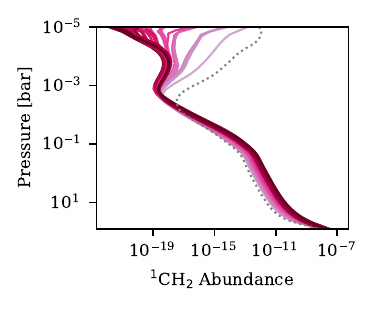}{0.5\columnwidth}{}
\fig{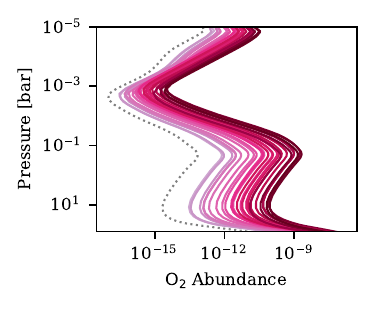}{0.5\columnwidth}{}}
\gridline{\fig{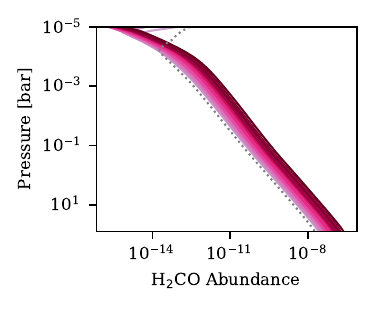}{0.5\columnwidth}{}
\fig{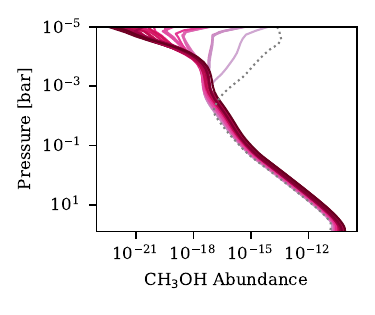}{0.5\columnwidth}{}
\fig{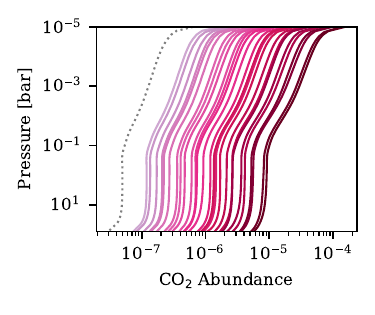}{0.5\columnwidth}{}}
\gridline{\fig{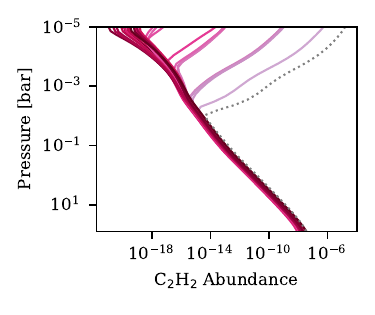}{0.5\columnwidth}{}
\fig{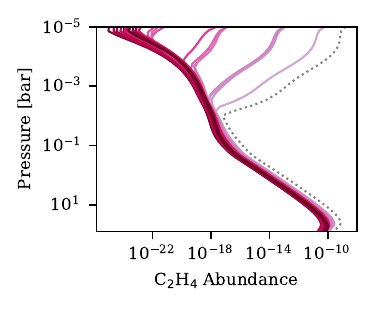}{0.5\columnwidth}{}
\fig{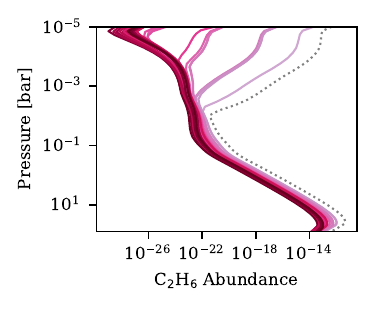}{0.5\columnwidth}{}}
\caption{Figure Set 5.1}
\end{figure*}
\begin{figure*}
\gridline{\fig{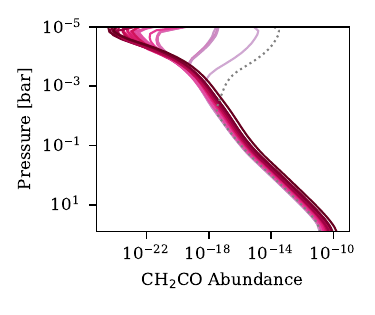}{0.5\columnwidth}{}
\fig{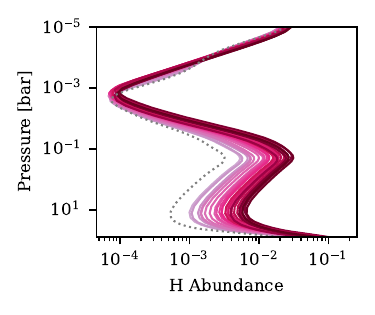}{0.5\columnwidth}{}
\fig{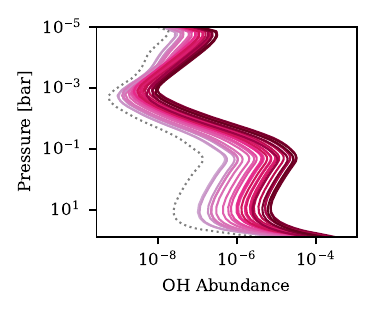}{0.5\columnwidth}{}}
\gridline{\fig{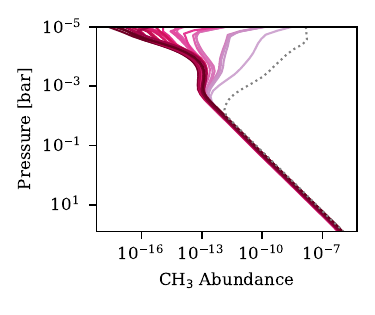}{0.5\columnwidth}{}
\fig{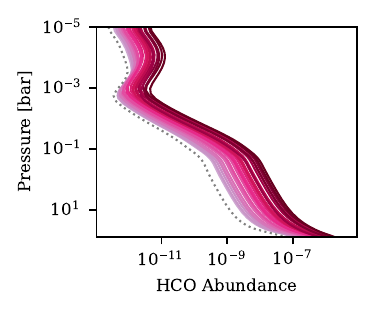}{0.5\columnwidth}{}
\fig{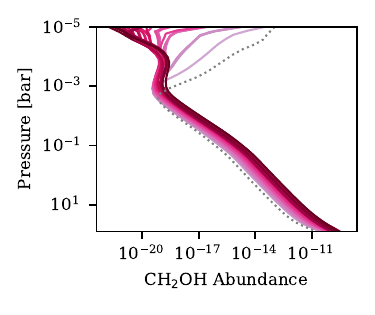}{0.5\columnwidth}{}}
\gridline{\fig{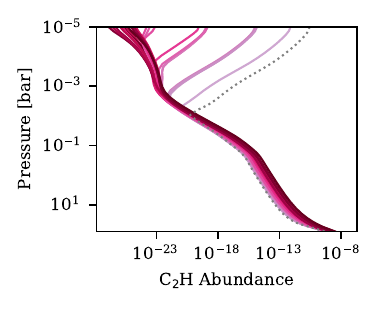}{0.5\columnwidth}{}
\fig{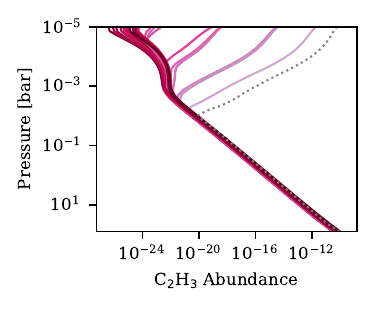}{0.5\columnwidth}{}
\fig{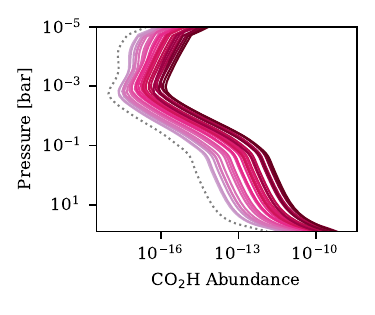}{0.5\columnwidth}{}}
\gridline{\fig{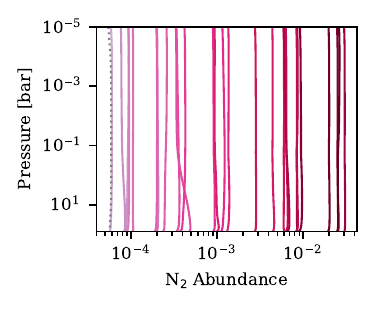}{0.5\columnwidth}{}
\fig{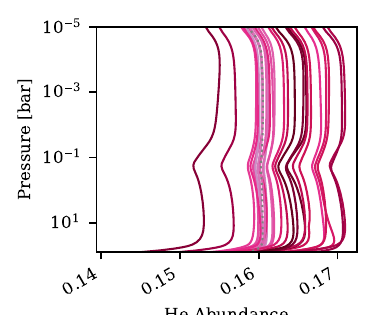}{0.5\columnwidth}{}
\fig{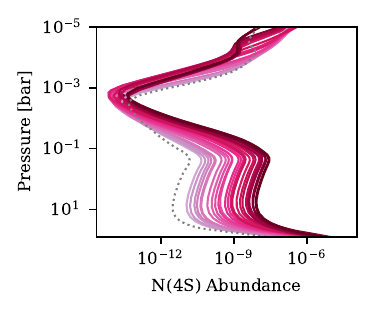}{0.5\columnwidth}{}}
\gridline{\fig{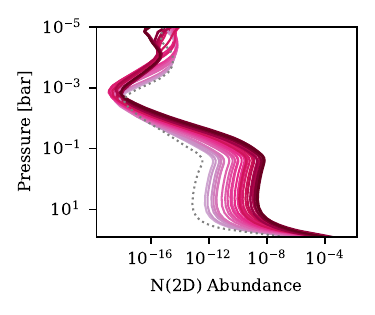}{0.5\columnwidth}{}
\fig{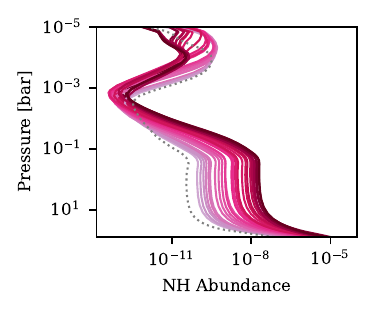}{0.5\columnwidth}{}
\fig{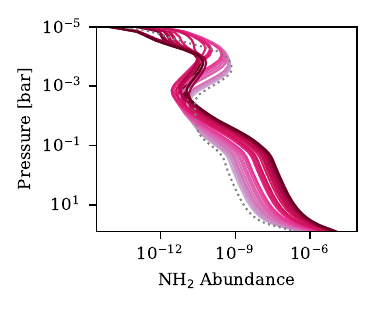}{0.5\columnwidth}{}}
\caption{Figure Set 5.2}
\end{figure*}
\begin{figure*}
\gridline{\fig{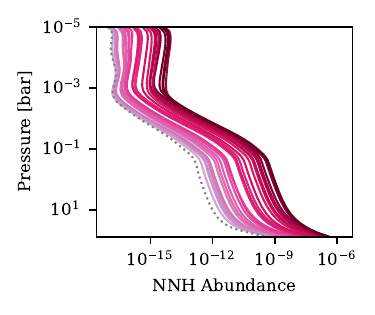}{0.5\columnwidth}{}
\fig{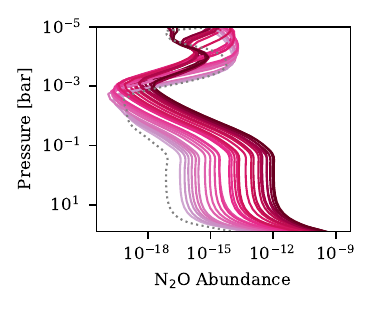}{0.5\columnwidth}{}
\fig{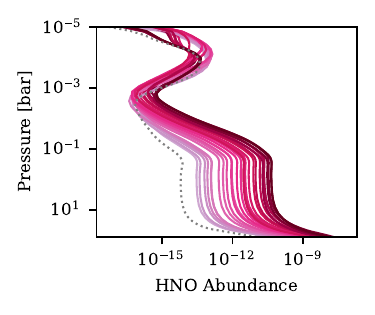}{0.5\columnwidth}{}}
\gridline{\fig{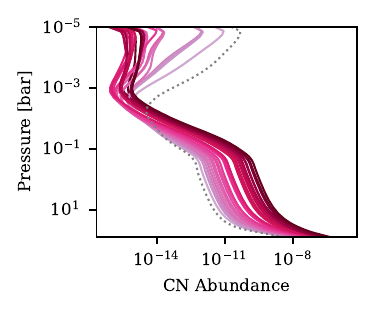}{0.5\columnwidth}{}
\fig{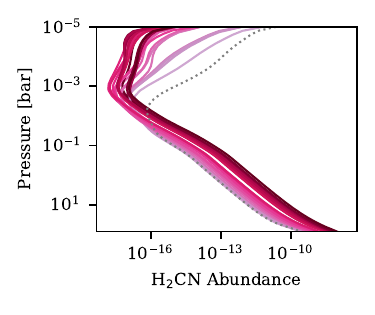}{0.5\columnwidth}{}
\fig{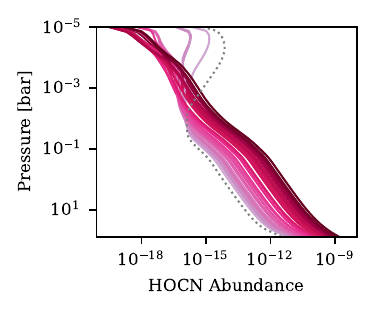}{0.5\columnwidth}{}}
\gridline{\fig{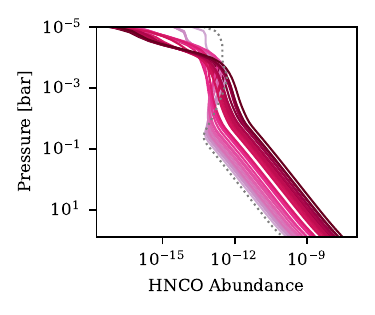}{0.5\columnwidth}{}
\fig{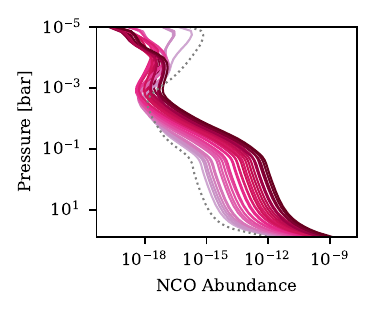}{0.5\columnwidth}{}
\fig{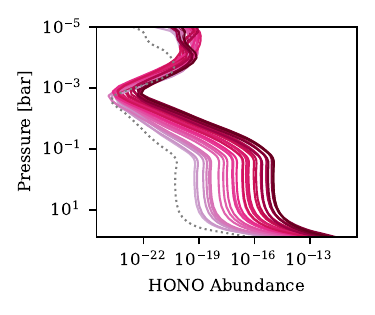}{0.5\columnwidth}{}}
\gridline{\fig{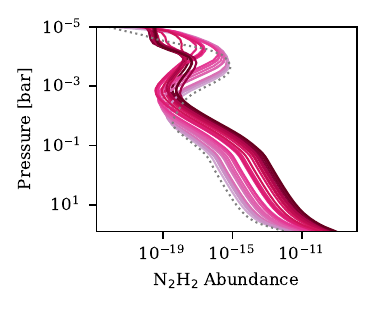}{0.5\columnwidth}{}
\fig{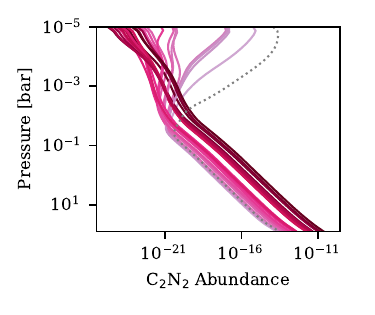}{0.5\columnwidth}{}
\fig{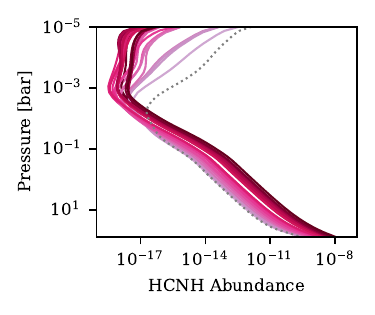}{0.5\columnwidth}{}}
\caption{Figure Set 5.3}
\end{figure*}
\begin{figure*}
\gridline{\fig{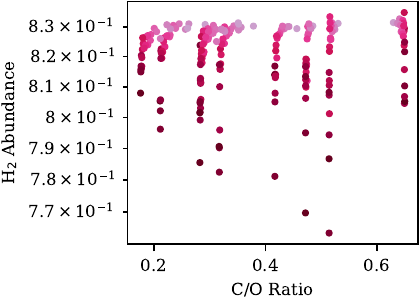}{0.5\columnwidth}{}
\fig{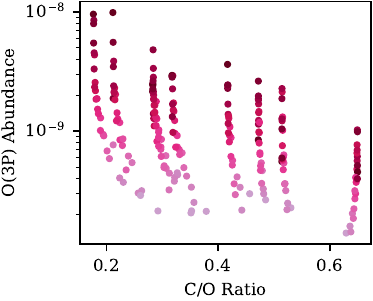}{0.5\columnwidth}{}
\fig{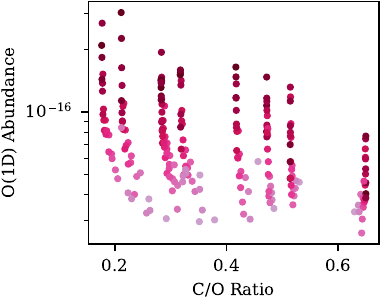}{0.5\columnwidth}{}}
\gridline{\fig{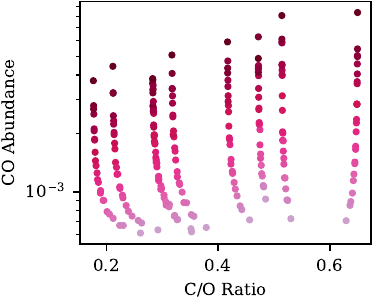}{0.5\columnwidth}{}
\fig{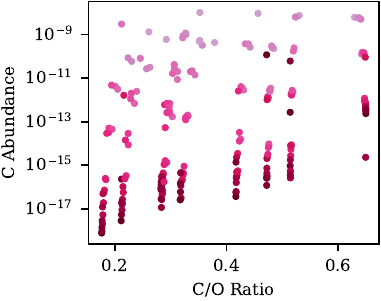}{0.5\columnwidth}{}
\fig{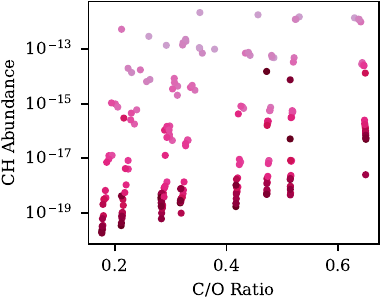}{0.5\columnwidth}{}}
\gridline{\fig{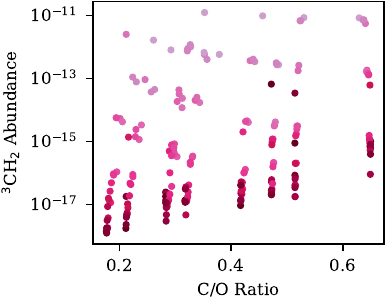}{0.5\columnwidth}{}
\fig{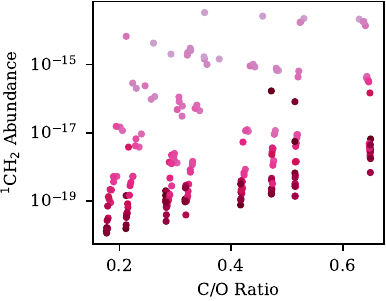}{0.5\columnwidth}{}
\fig{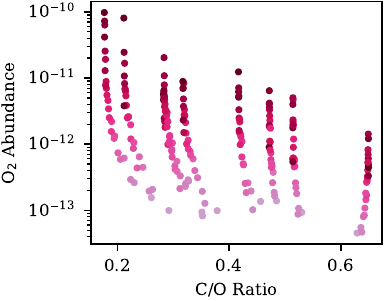}{0.5\columnwidth}{}}
\gridline{\fig{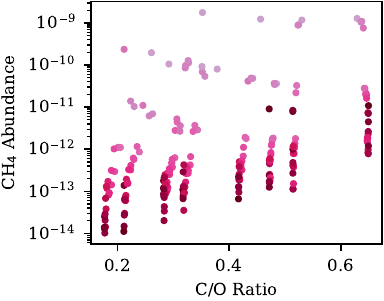}{0.5\columnwidth}{}
\fig{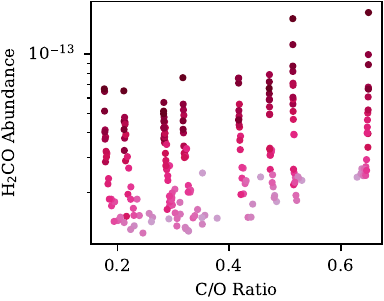}{0.5\columnwidth}{}
\fig{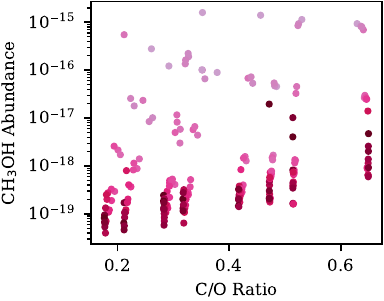}{0.5\columnwidth}{}}
\gridline{\fig{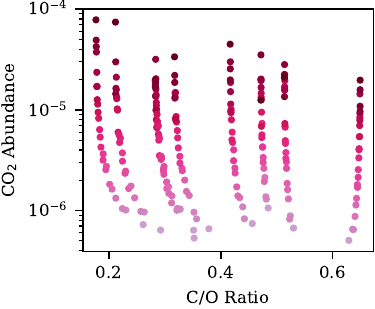}{0.5\columnwidth}{}
\fig{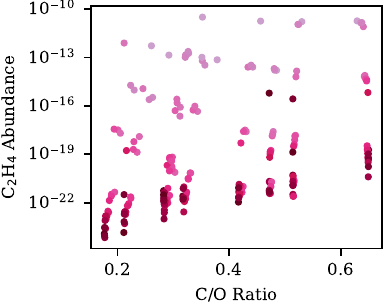}{0.5\columnwidth}{}
\fig{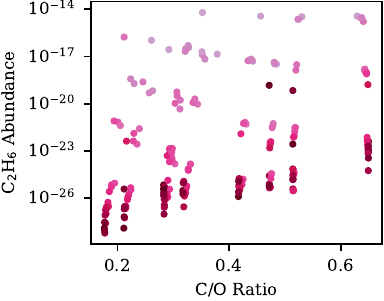}{0.5\columnwidth}{}}
\caption{Figure Set 6.1}
\end{figure*}
\begin{figure*}
\gridline{\fig{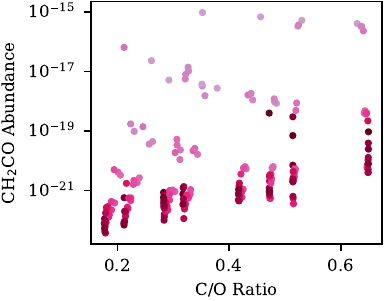}{0.5\columnwidth}{}
\fig{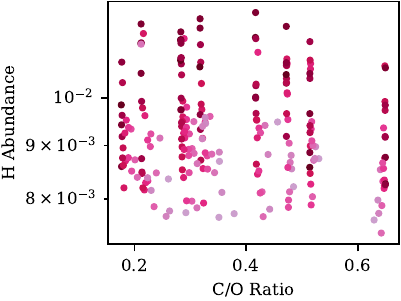}{0.5\columnwidth}{}
\fig{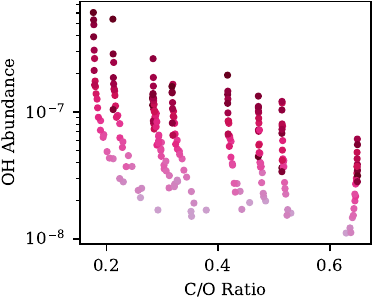}{0.5\columnwidth}{}}
\gridline{\fig{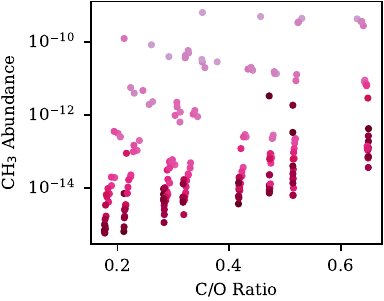}{0.5\columnwidth}{}
\fig{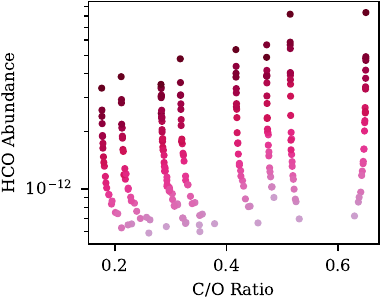}{0.5\columnwidth}{}
\fig{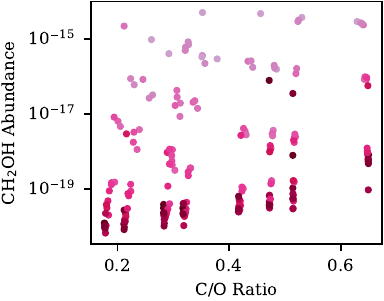}{0.5\columnwidth}{}}
\gridline{\fig{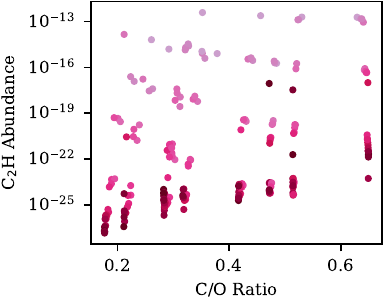}{0.5\columnwidth}{}
\fig{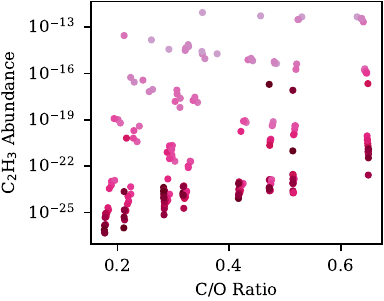}{0.5\columnwidth}{}
\fig{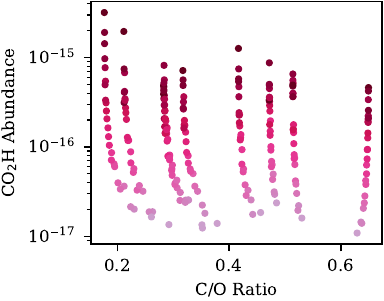}{0.5\columnwidth}{}}
\gridline{\fig{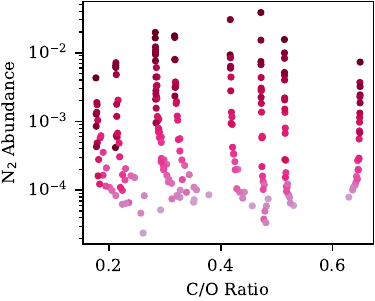}{0.5\columnwidth}{}
\fig{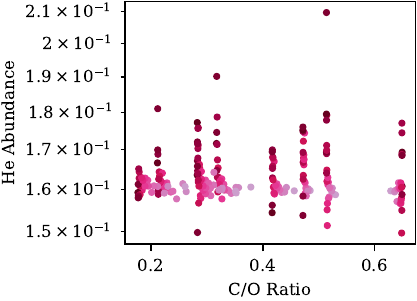}{0.5\columnwidth}{}
\fig{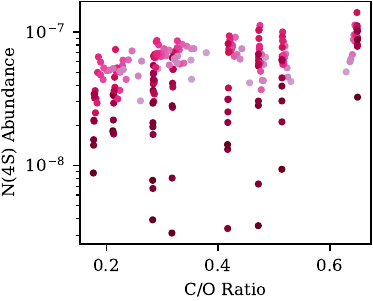}{0.5\columnwidth}{}}
\gridline{\fig{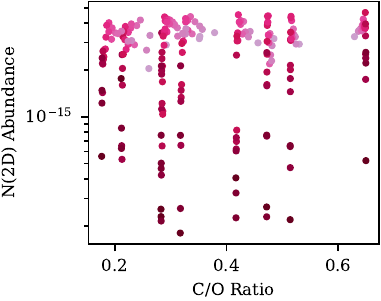}{0.5\columnwidth}{}
\fig{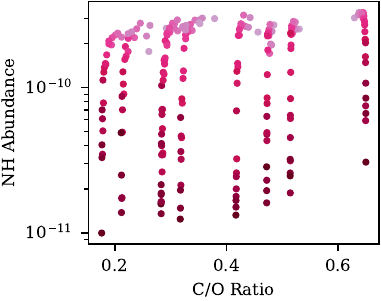}{0.5\columnwidth}{}
\fig{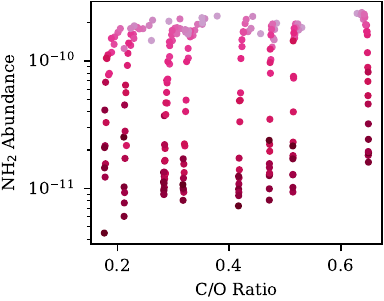}{0.5\columnwidth}{}}
\caption{Figure Set 6.2}
\end{figure*}
\begin{figure*}
\gridline{\fig{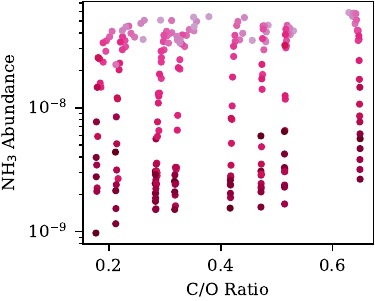}{0.5\columnwidth}{}
\fig{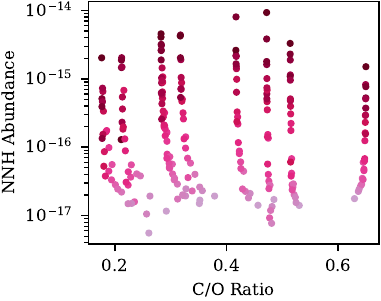}{0.5\columnwidth}{}
\fig{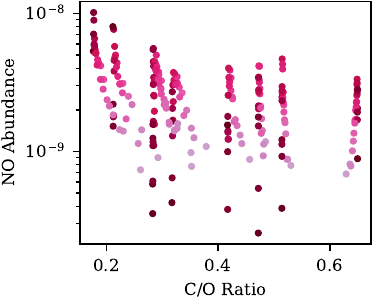}{0.5\columnwidth}{}}
\gridline{\fig{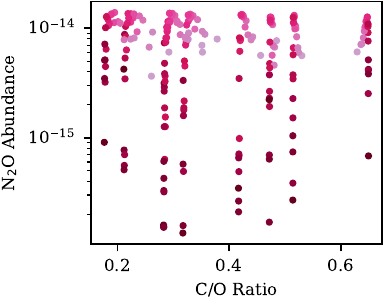}{0.5\columnwidth}{}
\fig{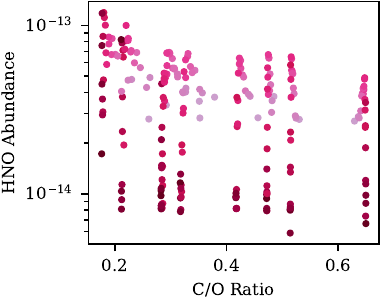}{0.5\columnwidth}{}
\fig{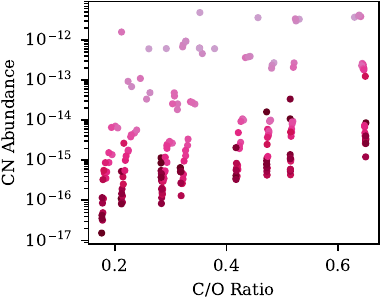}{0.5\columnwidth}{}}
\gridline{\fig{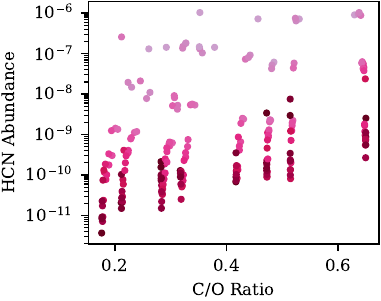}{0.5\columnwidth}{}
\fig{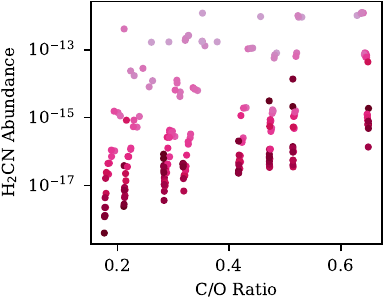}{0.5\columnwidth}{}
\fig{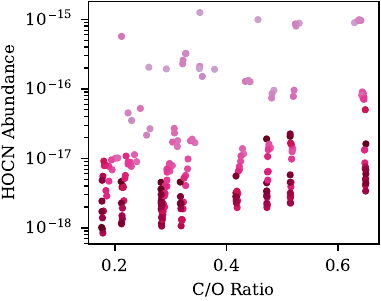}{0.5\columnwidth}{}}
\gridline{\fig{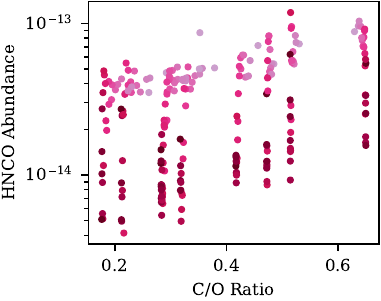}{0.5\columnwidth}{}
\fig{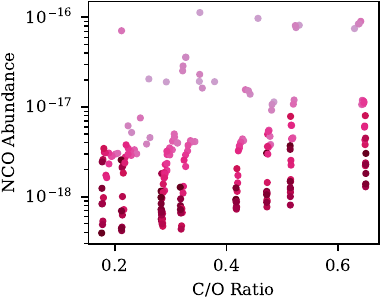}{0.5\columnwidth}{}
\fig{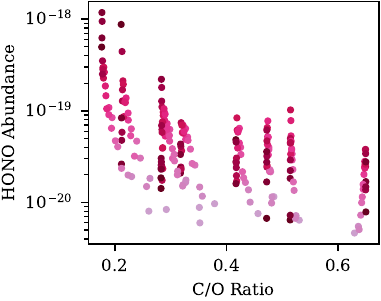}{0.5\columnwidth}{}}
\gridline{\fig{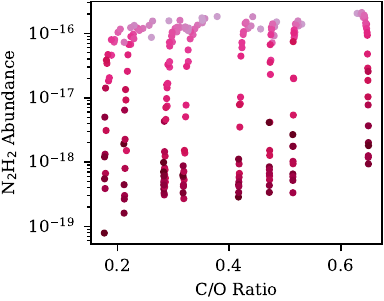}{0.5\columnwidth}{}
\fig{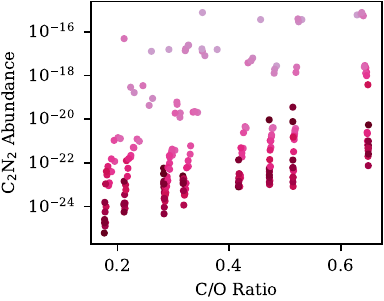}{0.5\columnwidth}{}
\fig{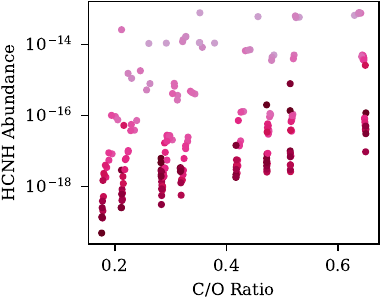}{0.5\columnwidth}{}
\fig{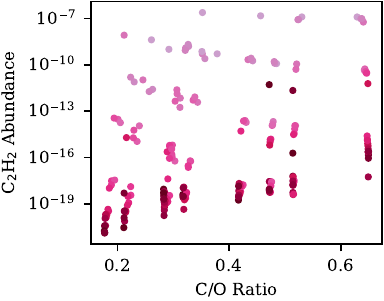}{0.5\columnwidth}{}}
\caption{Figure Set 6.3}
\end{figure*}

\begin{acknowledgements}
\nolinenumbers
F. Sainsbury-Martinez and C. Walsh would like to thank UK Research and Innovation for support under grant number MR/T040726/1. Additionally, C. Walsh would like to thank the University of Leeds and the Science and Technology Facilities Council for financial support (ST/X001016/1).\\
\end{acknowledgements}

\bibliography{papers}{}
\bibliographystyle{aasjournal}

\end{document}